%% file: sample-manuscript.tex
\RequirePackage{rotating}
\documentclass[acmsmall]{acmart}
\usepackage{booktabs}
\usepackage{adjustbox}
\usepackage{bm}
\usepackage{enumitem}
\usepackage{multirow}
\usepackage[ruled,linesnumbered]{algorithm2e}
\usepackage{colortbl}
\usepackage{balance}
\usepackage{rotating}

\DeclareMathOperator*{\argmax}{arg\,max}

\usepackage{pgfplots}
\usetikzlibrary{pgfplots.groupplots,intersections,pgfplots.fillbetween}
\pgfplotsset{compat=1.3}
\usepackage{scalefnt} 
\definecolor{color1}{RGB}{145,30,180}
\definecolor{color2}{RGB}{245,130,48}
\definecolor{color3}{RGB}{230,25,75}

\AtBeginDocument{%
  \providecommand\BibTeX{{%
    \normalfont B\kern-0.5em{\scshape i\kern-0.25em b}\kern-0.8em\TeX}}}

\setcopyright{acmlicensed}
\acmJournal{TOIS}
\acmYear{2025} \acmVolume{1} \acmNumber{1} \acmArticle{1} \acmMonth{1}\acmDOI{10.1145/3760401}

\begin{document}

%\title{Planning-Assessment-Interaction: Graph-enhanced Reinforcement Learning for Goal-oriented Intelligent Tutoring Systems}
\title{Towards Goal-oriented Intelligent Tutoring Systems in Online Education}

\author{Yang Deng}
\authornote{Both authors contributed equally to this research.}
\email{ydeng@smu.edu.sg}
\affiliation{%
  \institution{Singapore Management University}
  \country{Singapore}
}

\author{Zifeng Ren}
\authornotemark[1]
\email{renzifeng@u.nus.edu}
\affiliation{%
  \institution{National University of Singapore}
  \country{Singapore}
}

\author{An Zhang}
\authornote{Corresponding author.}
\affiliation{%
  \institution{University of Science and Technology of China}
  \country{China}}
\email{an.zhang3.14@gmail.com}

\author{Tat-Seng Chua}
\affiliation{%
  \institution{National University of Singapore}
  \country{Singapore}}
\email{chuats@comp.nus.edu.sg}

\renewcommand{\shortauthors}{Deng and Ren, et al.}

\begin{abstract}
Interactive Intelligent Tutoring Systems (ITSs) enhance the learning experience in online education by fostering effective learning through interactive problem-solving. However, many current ITS models do not fully incorporate proactive engagement strategies that optimize educational resources through thoughtful planning and assessment. In this work, we propose a novel and practical task of Goal-oriented Intelligent Tutoring Systems (GITS), designed to help students achieve proficiency in specific concepts through a tailored sequence of exercises and evaluations. We introduce a novel graph-based reinforcement learning framework, named Planning-Assessment-Interaction (\texttt{PAI}), to tackle the challenges of goal-oriented policy learning within GITS. This framework utilizes cognitive structure information to refine state representation and guide the selection of subsequent actions, whether that involves presenting an exercise or conducting an assessment. Additionally, \texttt{PAI} employs a cognitive diagnosis model that dynamically updates to predict student reactions to exercises and assessments. We construct three benchmark datasets covering different subjects to facilitate offline GITS research. Experimental results validate \texttt{PAI}'s effectiveness and efficiency, and we present comprehensive analyses of its performance with different student types, highlighting the unique challenges presented by this task.

%\noindent\textbf{Relevance Statement}: Online education (\textit{e.g.}, Massive Courses) is a popular and important web application for facilitating more accessible and inclusive interactions between the Web and society. 
\end{abstract}




\received{31 March 2024}

\maketitle

\section{Introduction}
Intelligent tutoring systems (ITSs) \cite{polson2013foundations}, which aim to provide personalized and effective instructional support to students, have gained increasing importance due to the growing demand for adaptive and accessible education in the society, especially in remote or online learning environments. They are applied in a wide range of web applications, such as MOOCs (Massive Open Online Courses) and various mobile learning apps, under the context from K-12 to higher education. 
Traditional ITSs often offer static and predefined content, which lack the dynamic interactivity and adaptability. 
Recent studies develop interactive ITSs \cite{cts-survey23,cts-survey21} that can provide real-time feedback \cite{tutor-llm,educhat}, engage in natural conversations \cite{cima,lrec22-talkmove,eacl23-tutor,eacl23fds-tutor}, and customize their teaching content based on individual student needs \cite{kdd19-ktsim,ktsim-learningpath,www21-learningpath}.  
The advent of large language models (LLMs) further empowers interactive ITSs with exceptional capabilities on natural language interactions \cite{llm-aied23,tutor-llm,educhat}. 
However, these studies mainly focus on the \textbf{reactive} engagement \cite{mallik2023proactive} of the interactive ITSs - to ensure that students acquire the necessary knowledge and to address questions raised by students during the interactions.  
While the \textbf{proactive} engagement \cite{mallik2023proactive} is often overlooked in the design of current interactive ITSs - to design and curate an optimal use of resources for achieving specific pedagogical goals, which requires the capabilities of \textit{planning} and \textit{assessment}.

\begin{figure}[t]
   \centering  
   \includegraphics[width=0.8\textwidth]{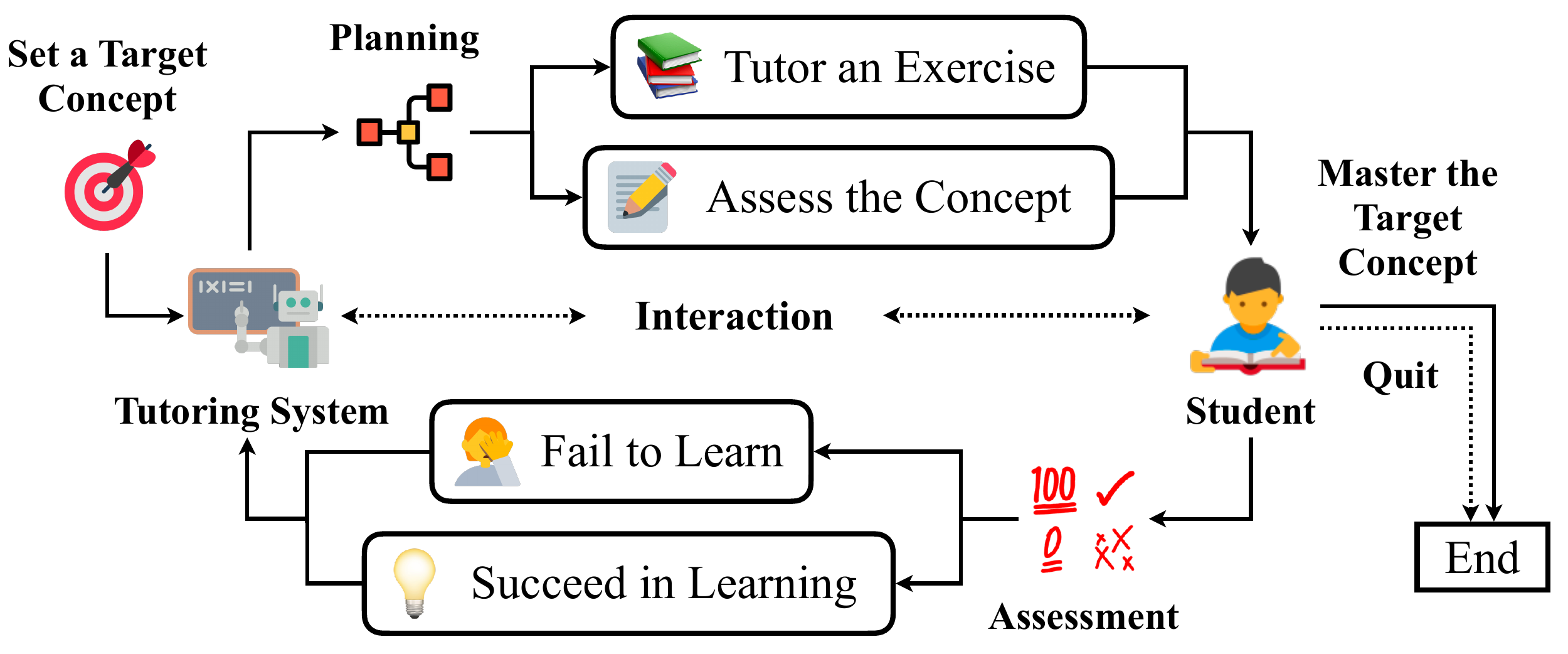} 
   \caption{The workflow of GITS. The goal of the tutoring system is to educate the student a specific target concept through a multi-turn interaction session. Specifically, the planning of GITS involves two types of actions, including 1) \textit{Tutor an Exercise} aims to tutor the student to comprehend an exercise for improving their understandings of the target concept. Afterwards, the system will decide the next action. 2) \textit{Assess the Concept} aims to assess the student regarding their mastery of the target concept. If the student fails to pass the assessment, the system will decide the next action. The interaction terminates if the student masters the target concept or quits the interactive learning session.} 
   \label{fig:problem}
\end{figure}  

Inspired by the remarkable success of goal-oriented interactive systems \cite{acl23-goal,recsys23-mm-crs,tois25-survey,sigir24-perspectives,iclr24-tutor} that can proactively guide the human-computer interaction toward predefined objectives, we introduce a new task, called Goal-oriented Intelligent Tutoring Systems (GITS), to investigate the proactive engagement in ITSs. As the workflow illustrated in Figure \ref{fig:problem}, an ITS engages in interactions with students, delivering a tailored sequence of exercises with a specific pedagogical goal that is to facilitate and accelerate the mastery of a predefined target concept by the student. Unlike those reactive ITSs, which may focus on individual exercises in isolation, GITS provides a cohesive and strategic learning experience, aligning closely with the student's long-term educational objectives. Two fundamental roles of proactive engagement in GITS are to determine:
\begin{itemize}
    \item[1)] \textit{What kinds of knowledge to be presented to students?} The ITS needs to determine which exercise to teach the student with two basic criteria: (i) The student can comprehend this exercise without losing their learning interests, ensuring that the exercise aligns with their current knowledge level – not too difficult or too elementary for their comprehension \cite{aaai20-truelearn}; and (ii) The mastery of the goal concept can benefit from comprehending this exercise. 
    \item[2)] \textit{When to assess students' mastery degree?} With sufficient certainty, the ITS should assess the student's mastery of the target concept. In contrast, assessing at the wrong moment can significantly impact their engagement and interest in learning. 
\end{itemize}
At each interaction turn, the ITS can choose to either \textit{tutor} the student with an exercise for improving their understandings of the target concept, or \textit{assess} the student regarding their mastery of the target concept. 
In return, the student may either succeed in learning the exercise/concept or fail to comprehend the exercise/concept. 
The interaction session will be terminated upon the student's successful mastery of the designated target concept or if, regrettably, they decide to discontinue the learning process. During a session, the ITS may switch between the above actions multiple times, with the goal of facilitating the student's mastery of the target concept while minimizing the overall number of interactions. 

In recent years, researchers have proposed various adaptive learning approaches \cite{cikm19-exercise,edm19-exercise,kdd19-ktsim} to personalize the learning path for improving the overall knowledge level of each individual student.  
However, these adaptive learning approaches encounter several challenges when addressing the GITS problem. 
(1) They primarily recommend a sequence of exercises to maximize the student's learning gain within a fixed number of interactions, but lack an effective and efficient plan for guiding students towards achieving a specific long-term educational goals, \textit{i.e.}, the target concept in GITS.   
(2) These methods only measure whether students can correctly respond to the exercise, overlooking the importance of assessing the mastery level of the underlying target concept within GITS. 
(3) Many of them rely on offline historical data to construct ITSs. This offline learning paradigm, rooted in static historical data, potentially misaligns with the dynamic nature of the online user learning process within interactive settings. 

To tackle these challenges, we propose a novel framework, named Planning-Assessment-Interaction (\texttt{PAI}), for the goal-oriented policy learning in GITS. In specific, we formulate the tutoring policy learning in GITS as a Markov Decision Process problem that can be optimized by reinforcement learning (RL), regarding the mastery of the target concept as the long-term goal. 
Firstly, we harness cognitive structure information, encompassing cognitive graphs to enhance state representation learning and prerequisite relations for refining action selection strategies. 
These elements work in tandem to enhance the goal-oriented policy planning for proactively achieving the pedagogical goal. 
Secondly, we implement a dynamically updated cognitive diagnosis model that simulates real-time student responses to exercises and concepts. This simulation accommodates diverse types of students by varying difficulty levels, learning patience, and learning speeds, facilitating research in online education with diversity, equity, and inclusion. 
Overall, we employ a graph-based RL algorithm to optimize the goal-oriented policy learning problem, with the aim to achieve the designated goal effectively and efficiently. 

To sum up, the main contributions of this work are as follows: 
\begin{itemize}
    \item We comprehensively consider a goal-oriented intelligent tutoring system (GITS) scenario that is a practical application in online education, highlighting the importance of researching into the designs of proactive engagement in ITSs. 
    \item We propose a novel RL-based framework, namely Planning-Assessment-Interaction (\texttt{PAI}), to leverage both cognitive structure information and cognitive diagnosis techniques for the goal-oriented policy learning in GITS.  
    \item We build three GITS datasets simulating teacher-student interactions to enable offline academic research.  
    Experimental results demonstrate the effectiveness and efficiency of \texttt{PAI} and extensive analyses showcase the challenges presented in this task. \footnote{Code and data will be released via \url{https://github.com/Sky-Wanderer/Towards-Goal-oriented-Intelligent-Tutoring-Systems-in-Online-Education}.}
\end{itemize}

\section{Related Works}
This work is closely related to the following research areas: 
\subsection{Interactive Intelligent Tutoring Systems}
As an advanced form of intelligent tutoring systems (ITSs), interactive ITSs has been extensively investigated as educational dialogue systems \cite{chi19-quizbot,cts-survey21,eacl23-tutor}, as it can interatively provide adaptive instructions and real-time feedbac, so that students can learn more efficiently and more engaged in study.
Most existing studies focus on learning the pedagogical strategies to teach the students of the given exercises \cite{cima,lrec22-talkmove,tois-2023}.
For example, \citet{cima} collect tutoring dialogues dataset reflecting pedagogical strategies through role-playing crowdworkers. The dataset highlights reduced student turn-taking and tutors adhering to educational conversational norms, aiding in training models for generating tutoring utterances. \citet{lrec22-talkmove} introduce the TalkMoves dataset, enriched with annotations from K-12 mathematics lessons which emphasizes that good tutoring dialogue strategy can promote equitable student participation and explicit thinking.
Some studies focused on generating high-quality responses in the tutoring dialogues. \citet{eacl23fds-tutor} introduce a unified framework for conversational tutoring systems (CTSs), jointly predicting teaching strategies and generating tutor responses which addresses the challenge of engaging students with diverse teaching strategies, enhancing realism and learning outcomes.
\citet{tois-2023} enhance automated classification of instructional strategies in online tutorials by incorporating contextual information and active learning methods, which improves machine learning models and reduces the need for manual data annotation. 
\citet{tois24-its} introduce a heterogeneous evolution network (HEN) for learning the representations of entities and relations of the educational concepts for ITSs. 

Latest studies \cite{llm-aied23,tutor-llm,educhat,hci-2024,acl24-qg} on interactive ITSs powered by LLMs have showcased the exceptional capabilities on natural language interactions. 
\cite{hci-2024} propose a personalized tutoring system, emphasizing diagnostic assessments, conversation-based tutoring with LLMs, and interaction analysis, which informs potential enhancements and invites HCI collaboration in personalized education technology.
However, most existing ITSs play a passive role in the interactive engagement with students, such as ensuring students' understanding of knowledge or addressing their questions. 
In this work, we investigate the proactive engagement in interactive ITSs \cite{mallik2023proactive}, which emphasizes resource optimization to strategize proactive tutoring through planning and assessment, instead of delving into content generation during the interaction.

\subsection{Adaptive Learning in Online Education} 
Adaptive learning \cite{adaptive-learning}, also called adaptive tutoring, is a method that utilizes personalized recommendation techniques to suggest learning materials, such as lectures or exercises, to meet the distinct requirements of each student. 
Early studies adopt sequential recommendation methods to generate learning paths \cite{infsci18-learningpath,lak19-goal-course}. \citet{infsci18-learningpath} introduce a RNN-based method for personalized course prerequisite inference, offering tailored course recommendations to students for desired achievement goals. \citet{lak19-goal-course} present a novel LSTM neural network model for a full-path learning recommendation system, addressing challenges in personalized online education with clustered data analysis to improve learning path predictions and mitigate the cold-start problem in e-learning environments. 

Some studies also select the next exercise to tutor \cite{edm19-exercise,cikm19-exercise}. 
\citet{edm19-exercise} integrate course concepts and exercise-concept mappings, improving knowledge tracing and input features and used deep reinforcement learning for personalized math exercise recommendations. \citet{cikm19-exercise} use a flexible Q-Network for exercise selection, state learning with multi-faceted educational data, and the novel optimization of three educational objectives, enabling adaptive exercise recommendations.

However, these methods train recommendation models using static historical data, which limits their ability to optimize performance offline and may not fully align with the dynamic nature of user learning in reality. 
Another line of research mainly focuses on the online assessment of the student's knowledge state by recommending exercises \cite{sigir22-assess,aaai22-assess,cikm23-assess,sigir22-question,tois-tracing}. 
Recently, researchers improve the adaptive learning by applying pretraining techniques over heterogeneous learning elements \cite{pretrain-adaptive-learning,sigir23-cd}, employing RL techniques to learn from long-term rewards \cite{kdd19-ktsim,ktsim-learningpath,aaai23-learnpath,kdd23-exercise}, and leveraging prior structured knowledge \cite{www21-learningpath,ktsim-exercise,sigir21-rcd,sigir22-kt,tois-graph}. \citet{tois-graph} introduce DGEKT, a graph ensemble learning method that captures the heterogeneous exercise-concept associations and interaction transitions through dual graph structures. 
Moreover, they solely focus on the improvement of overall knowledge level of the student within a fixed number of exercises, but neglect the measurement of the tutoring efficiency and the student's learning interest as well as fail to make strategic plans for achieving designated goals. \citet{tois-tracing} develop two models, Knowledge Proficiency Tracing (KPT) and Exercise-correlated KPT (EKPT), that enhance student learning analysis by integrating Q-matrix, learning and forgetting curves, and exercise connectivity. KPT maps exercises and student proficiencies in a shared knowledge space, while EKPT further improves predictions by linking exercises.

\subsection{Goal-conditioned Reinforcement Learning}
In contrast to conventional RL approaches that rely solely on states or observations to learn policies, Goal-conditioned reinforcement learning (GCRL) \cite{ijcai22-gcrl} tackles complex RL problems by training an agent to make decisions based on diverse goals in addition to environmental cues. 
For example, GCRL has been widely introduced into interactive recommender systems \cite{nips19-irs-rl,wsdm2020-irs-rl,sigir20-graph-irl,tois-rec} and conversational recommender systems \cite{sigir21-unicorn,www22-crs,www23-crs-user,recsys23-mm-crs,wsdm-convrec,tois-recconv} due to its advantage of considering users' long-term feedback and capture users' dynamic preferences for generating accurate recommendations over time. 
\citet{tois-rec} combine offline RL with causal inference to mitigate filter bubbles by learning a causal user model for interest and overexposure, using counterfactual satisfaction for RL policy planning, and evaluating policies by cumulative user satisfaction in real settings.
\citet{tois-recconv} introduce a meta-reinforcement learning framework for conversational recommender systems, employs a dynamic, personalized knowledge graph and model-based learning to adapt recommendations based on user interactions and feedback. 
The objectives of these approaches typically are to learn an effective policy for determining the recommended items. However, it casts a new challenge on applying GCRL on GITS, since it  not only requires to  consider some prerequisite dependencies \cite{acl17-prerequisite,aaai18-prerequisite} that adds an additional layer of complexity for recommendation but also poses difficulties on user simulation that involve cognitive diagnosis \cite{aaai20-cdm,tois-tracing} for assessing the user's knowledge state.

\begin{table}[t]
%\fontsize{7}{9}\selectfont
\centering
  \caption{Notations.}
  \begin{tabular}{lp{10cm}}
    \toprule
	Notation & Definition  \\
    \midrule
    $\mathcal{U}$ & the student set\\
    $\mathcal{E}$ & the exercise set\\
    $\mathcal{C}$ & the concept set\\
    $O$ & the student-exercise matrix\\
    $Q$ & the concept-exercise matrix\\
    $P$ & the prerequisite adjacent matrix\\
    $c^*$ & the target concept\\
    $\mathcal{E}_c$ & the set of exercises that belong to the concept $c$ \\
    $\mathcal{E}_\text{cand}^{(t)}$ & the candidate exercise set at turn $t$ \\
    $\mathcal{E}_+^{(t)}$ & the set of appropriate exercises that have been previously tutored at turn $t$ \\
    $\mathcal{E}_-^{(t)}$ & the set of inappropriate exercises that have been previously tutored at turn $t$ \\
    $\mathcal{A}_t$ & the action space at turn $t$\\
    $a_t$ & the action taken at turn $t$\\
    $s_t$ & the state at turn $t$\\
    $r_t$ & the reward at turn $t$\\
    $f_t$ & the student response at turn $t$\\
    $l_t$ & the student's patience loss at turn $t$\\
    $w_e^{(t)}$ & the exercise score at turn $t$\\
    $w_c^{(t)}$ & the concept score at turn $t$\\
    $\rho_{u,e}$ & the probability of the student $u$ correctly responds to the exercise $e$\\
    $d_{u,c}$ & the estimated mastery level of the concept $c$ for the student $u$\\
    $T$ & the maximum number of interaction turn\\
    $\beta$ & the maximum patience loss of the student\\
    $\delta$ & the threshold score of passing the examination\\
    $\lambda_+$ & the upper threshold of the appropriate difficulty level of the exercise\\
    $\lambda_-$ & the lower threshold of the appropriate difficulty level of the exercise\\
    $\alpha$ & the learning rate of the dynamical update\\
  \bottomrule
\end{tabular}
\label{tab:notation}
\end{table}

\section{Problem Definition}\label{sec:problem}
In online education~\cite{learning-goal}, the learning goals are typically defined as the specific domain concepts that are supposed to be mastered by the student. To achieve the learning goals, the learning path may involve prerequisite concept hierarchy~\cite{acl17-prerequisite,aaai18-prerequisite} or related learning materials, such as exercises~\cite{edm19-exercise,cikm19-exercise}. 
Since the mastery level of concepts depends on the teaching materials~\cite{ai90-its,csur23-ktsurvey}, the system can only facilitate mastery of specific concepts by engaging students with  related exercises, rather than simply providing direct instructions on the concept itself. 
Accordingly, we denote a designated target concept as a learning goal. 
To achieve this goal, the ITS can either assess the student's mastery of the target concept or tutor the student to comprehend related exercises. 

We introduce the notations used to formalize the problem of \textit{Goal-oriented Intelligent Tutoring System} (GITS). 
$u\in\mathcal{U}$ denotes a student $u$ from the student set $\mathcal{U}$. 
$c\in\mathcal{C}$ denotes a concept $c$ from the concept set $\mathcal{C}$. 
$e\in\mathcal{E}$ denotes a exercise $e$ from the exercise set $\mathcal{E}$. 
An student-exercise matrix $O\in\{-1,0,1\}^{|\mathcal{U}|\times|\mathcal{E}|}$ represents the past interactions between students and exercises, where -1 and 1 indicate the student incorrectly and correctly answers the exercise respectively while 0 indicates the student has yet answered the exercise. 
An concept-exercise matrix $Q\in\{0,1\}^{|\mathcal{E}|\times|\mathcal{C}|}$ represents the association between each exercise and concept. 
A prerequisite adjacent matrix $P\in\{0,1\}^{|\mathcal{C}|\times|\mathcal{C}|}$ denotes the prerequisite relation among concepts. 
As show in Figure~\ref{fig:problem}, the system is assigned with a designated target concept $c^*$ to start the interactive learning process. 
In each turn $t$, the ITS needs to choose an action: \textit{tutor} or \textit{assess}: 
\begin{itemize}
    \item If the action is \textit{tutor}, we denote the selected exercise for tutoring as $e\in\mathcal{E}$. Then the system can initiate a tutoring sub-session for teaching the student about the exercise. 
    After tutoring, if the student correctly comprehend the exercise, $O_{u,e}$ will be set to 1. 
    If not, $O_{u,e}$ will be set to -1. 
    At the same time, the student will lose certain learning patience, which is related to the difficulty of comprehending this exercise. 
    \item If the action is \textit{assess}, we let the student to conduct an examination containing exercises that are related to the target concept $c^*$. If the student passes the exam, we regard that the student has mastered the target concept, which means this learning session succeeds and can be terminated. If the student fails in the exam, the system moves to the next round and the student also loses certain learning patience. 
\end{itemize}
The whole process naturally forms an interaction loop, where the ITS may assess the mastery level of the student on %several prerequisite concepts 
the target concept 
or tutor the student to comprehend several exercises. 
The interaction terminates if the student masters the target concept or leaves the interactive learning process due to their impatience. 
The objective of GITS is to enable the student to master the target concept within as few rounds of interactions as possible. The summary of notations used in this work is presented in Table \ref{tab:notation}. 

\begin{figure}[t]
   \centering  
   \includegraphics[width=0.8\textwidth]{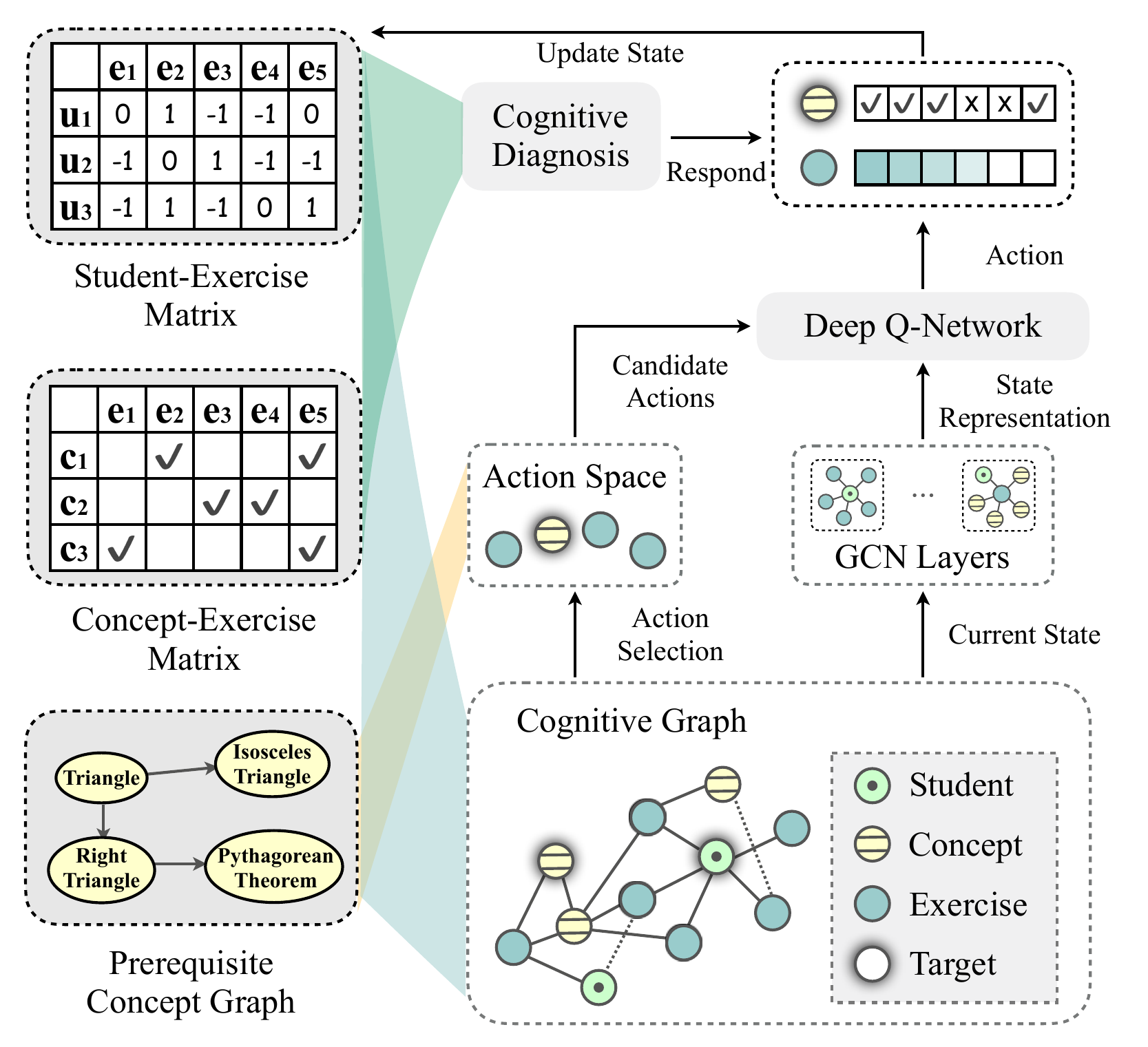} 
   \caption{The overview of the Planning-Assessment-Interaction (\texttt{PAI}) framework. \texttt{PAI} consists of three main components. 1) The Planning component (\S \ref{sec:planning}) applies cognitive graph structure for state representation learning and prerequisite structure for action selection. 2) The Assessment component (\S \ref{sec:assessment}) simulates the student rewards via dynamically updated cognitive diagnosis. 3) The Interaction component (\S \ref{sec:interaction}) performs the reinforcement learning with a Deep Q-Network.} 
   \label{fig:method}
\end{figure}

\section{Method}
We formulate the tutoring policy learning in GITS as a Markov Decision Process (MDP) problem, where the system aims to educate the student a specific target concept through a multi-turn interaction session. 
Given the state $s_t$ at the current timestep $t$, the ITS selects an action according to its policy $a_t\sim \pi(a|s_t)$, either assessing the student's mastery of the target concept or tutoring the student with an exercise.  
In return, the system receives a reward $r_t=\mathcal{R}(s_t,a_t)$ from the student feedback. 
This process repeats until the student masters the target concept or quits the interaction due to their impatience (\textit{e.g.}, reach the maximum interaction turns $T$). 
The objective of GITS is to learn the policy $\pi^*$ to maximize the expected cumulative rewards over the observed interactions: 
\begin{equation}
    \pi^* = \arg \max_{\pi\in\Pi} \left[\sum_{t=0}^T \mathcal{R}(s_t,a_t)\right].
\end{equation}
The overview of the proposed method, named Planning-Assessment-Interaction (\texttt{PAI}), is depicted in Figure \ref{fig:method}.

\subsection{Planning via Cognitive Structure}\label{sec:planning}
\subsubsection{\textbf{Graph-based State Representation Learning}}\label{sec:state}
We combine the student-exercise matrix $O$, the concept-exercise matrix $Q$, and the concept prerequisite matrix $P$ as a unified cognitive graph $\mathcal{G}$. To make use of the interrelationships among students, exercises, and concepts, we initially employ graph-based pre-training approaches~\cite{transe,transh} to acquire node embeddings $\{h\}$ for all nodes within the full graph $\mathcal{G}$.
Given a sample concerning the student $u$ and the target concept $c^*$, we denote the subgraph as $\mathcal{G}_{u,c^*}=(\mathcal{N},A)$, where $\mathcal{N}$ and $A$ denote the node set and the adjacent matrix: 
\begin{equation}
\mathcal{N} = \{u\}\cup \mathcal{E}_{c^*} \cup  \mathcal{C}
\end{equation} 
\begin{equation}
    \bm{A}_{i,j} =
  \begin{cases}
    O_{i,j},     & \quad \text{if } n_i \in \mathcal{U}, n_j \in \mathcal{E} \\
    Q_{i,j},     & \quad \text{if } n_i \in \mathcal{E}, n_j \in \mathcal{C} \\
    P_{i,j},     & \quad \text{if } n_i \in \mathcal{C}, n_j \in \mathcal{C} \\
    0,       & \quad \text{otherwise}
  \end{cases} 
\label{eq:a}
\end{equation}
where $\mathcal{E}_{c^*}$ is the set of exercises related to the target concept $c^*$. 

We denote the interaction history at the timestep $t$ as $\mathcal{H}^t = \{(a_i,f_i)\}_{i=1}^t$, where $f_i$ is the student feedback to the agent's action $a_i$. The state at the timestep $t$ is represented by $s_t = [\mathcal{G}_{u,c^*},\mathcal{H}^t]$, where the subgraph $\mathcal{G}_{u,c^*}$ is updated with the conversation history $\mathcal{H}^t$, \textit{i.e.}, $\mathcal{G}_{u,c^*}^t=(\mathcal{N},A^t)$:
\begin{equation}\label{eq:transition}
    A^t_{u,a_i} = f_i, \quad \text{ for } (a_i,f_i) \in \mathcal{H}^t  \text{ and } a_i \in \mathcal{E}.
\end{equation}

The current state $s_t$ will transition to the next state $s_{t+1}$ after the student finishes the learning with the feedback $f_t$ to the action $a_t$, where $f_t=1$ if the student succeeds in learning, otherwise $f_t=-1$. Then $\mathcal{H}^{t+1}=\mathcal{H}^t\cup (a_t,f_t)$ and $\mathcal{G}_{u,c^*}^{t+1}$ will be updated via Eq.(\ref{eq:transition}).  

To fully exploit the correlation information among students, items, and attributes within the interconnected graph, we utilize a graph convolutional network (GCN)~\cite{gcn} to enhance the node representations by incorporating structural and relational knowledge. The representations of node $n_i$ in the $(l+1)$-th layer can be calculated as follows:
\begin{equation}
    h_i^{(l+1)} = \text{ReLU}\left(\sum_{j\in \mathcal{N}_i}\bm{\Lambda}_{i,j}\bm{W}_{l}h_j^{(l)} + \bm{B}_lh_i^{(l)}\right),
\end{equation}
where $\mathcal{N}_i$ denotes the neighboring indices of node $n_i$, $\bm{W}_{l}$ and $\bm{B}_{l}$ are trainable parameters representing the transformation from neighboring nodes and node $n_i$ itself, and $\bm{\Lambda}$ is a normalization adjacent matrix as $\bm{\Lambda} = \bm{D}^{-\frac{1}{2}}\bm{A}\bm{D}^{-\frac{1}{2}}$ with $\bm{D}_{ii} =\sum_j \bm{A}_{i,j} $. 

The learned representations of the student $u$ and the target concept $c^*$ are passed over a mean pooling layer to obtain the state representation of $s_t$:
\begin{equation}\label{state_rep}
    f_{\theta_S}(s_t) = \mathbf{MeanPooling}([h^{(L)}_u;h^{(L)}_{c^*}]),
\end{equation}
where $\theta_S$ is the set of all network parameters for state representation learning, and $L$ is the number of layers in GCN. 
\subsubsection{\textbf{Prerequisite-guided Action Selection}}\label{sec:action_selection}
According to the current state $s_t$, the agent selects an action $a_t$ from the candidate action space $\mathcal{A}_t$, including the target concept $c^*$ and the candidate exercise set $\mathcal{E}_\text{cand}^{(t)}$. 
The candidate action space can be set to the whole action space, including all exercises and the target concept. However, this is impractical under a large action space in some applications, which will significantly harm the performance and efficiency with limited online interaction data. 
Furthermore, the learning concepts exhibit inherent cognitive structural characteristics, such as prerequisite relationships, which can be utilized not only to ensure logical and explainable decision-making but also to reduce the large search space of candidate actions. 
To this end, we design an action selection strategy to narrow down the action search space, based on the connectivity between the target concept $c^*$ and its predecessors $P_{c^*}$ in the prerequisite graph.  

Detailed process about the action selection strategy is presented in Algorithm \ref{algo-action-selection}.
At the turn $t$, we have the set of appropriate and inappropriate exercises previously tutored, \textit{i.e.}, $\mathcal{E}_+^{(t)}$ and $\mathcal{E}_-^{(t)}$. 
The goal is to obtain the candidate exercise set $\mathcal{E}_\text{cand}^{(t)}$ with $N$ candidate exercises. 
In specific, we first compute the exercise scores $w_e^{(t)}$ for each exercise $e$ in $\mathcal{E}_\text{cand}^{(t)}$ via Eq.(\ref{eq:exercise_score}). Then, based on the exercise scores $w_e^{(t)}$, we compute the concept scores $w_c^{(t)}$ for each concept $c \in P_{c^*}$ that is the prerequisite concept of the target concept $c^*$. Overall, we apply two levels of sorting for the exercises: 1) We prioritize the exercises related to the prerequisite concept with a higher concept score $w_c^{(t)}$; 2) Under the same prerequisite concept, we put the exercise with a higher exercise score $w_e^{(t)}$ into the candidate exercise set $\mathcal{E}_\text{cand}^{(t)}$ until the number of candidate exercises reaches $N$.

\textbf{Exercise Score:} In order to incorporate the user knowledge level as well as the  correlation with the previously tutored exercise, we first compute the exercise score based on the current state at the timestep $t$:
\begin{equation}
    w_e^{(t)} = \sigma(h_u^\top h_e + \sum_{e'\in \mathcal{E}_+^{(t)}} h_e^\top h_{e'} - \sum_{e'\in \mathcal{E}_-^{(t)}} h_e^\top h_{e'}),
    \label{eq:exercise_score}
\end{equation}
where $\mathcal{E}_+^{(t)}$ and $\mathcal{E}_-^{(t)}$ denote the sets of previously tutored exercises that is appropriate  and not appropriate (either too difficult or too easy) for the current state of the student to learn.

\textbf{Concept Score:} Furthermore, the expected exercise is supposed to be related to the prerequisite concept that can better eliminate the uncertainty of the target concept. 
Motivated by this, we adopt weighted entropy as the criteria to rank the set of exercises that is related to each prerequisite concept:
\begin{equation}
\begin{split}
    w_c^{(t)} &= -\text{prob}(c^{(t)}) \cdot \log (\text{prob}(c^{(t)})),\\
    \text{prob}(c^{(t)}) &= \sum_{e\in  \mathcal{E}_{c^*} \cap \mathcal{E}_c} w_e^{(t)}/\sum_{e\in \mathcal{E}_c} w_e^{(t)}.
\end{split}
\label{eq:concept_score}
\end{equation}

Overall, the prerequisite-guided action selection strategy select the top-$N$ exercises based on the exercise score in Eq.(\ref{eq:exercise_score}) from the set of exercises that belong to the prerequisite concept with the higher concept score in Eq.(\ref{eq:concept_score}). These top-$N$ exercises serve as the candidate exercise set $\mathcal{E}_\text{cand}^{(t)}$.

\begin{algorithm}[t]
\caption{Prerequisite-guided Action Selection}
\label{algo-action-selection}
\KwIn{The prerequisite graph $P$; the target concept $c^*$; the set of appropriate exercises previously tutored $\mathcal{E}_+^{(t)}$; the set of inappropriate exercises previously tutored $\mathcal{E}_-^{(t)}$; the number of candidate exercises $N$;  } 
\KwOut{The candidate exercise set $\mathcal{E}_\text{cand}^{(t)}$; }
Initialize $\mathcal{E}_\text{cand}^{(t)}=\varnothing$\;% $\mathcal{E}' = \mathcal{E}\backslash(\mathcal{E}_+^{(t)}\cup \mathcal{E}_-^{(t)})$\;
\For{$e\in \mathcal{E}$}{
Compute exercise score $w_e^{(t)}$ via Eq.(\ref{eq:exercise_score});
}
%\If{$\mathcal{C}_u^{(t)}=\varnothing$}
%{$\mathcal{C}_u^{(t)}=c_0$\tcp*{initialize with a void concept}} 
\For{$c\in P_{c^*}$}{
Compute prerequisite concept score $w_c^{(t)}$ via Eq.(\ref{eq:concept_score});
}
Sort $P_{c^*}$ by $w_c^{(t)}$\;
\For{$c\in P_{c^*}$}{
    Sort $\mathcal{E}_c$ by $w_e^{(t)}$\;
    \For{$e\in \mathcal{E}_c$}{
        $\mathcal{E}_\text{cand}^{(t)} = \mathcal{E}_\text{cand}^{(t)} \cup \{e\} $\;
        \If{$|\mathcal{E}_\text{cand}^{(t)}|=N$}
        {return $\mathcal{E}_\text{cand}^{(t)}$\;} 
    }
}
\end{algorithm}

\subsection{Assessment via Cognitive Diagnosis}\label{sec:assessment}
\subsubsection{\textbf{Reward}}
To align with the objective of GITS, we define five types of rewards: 1) $r_{c+}$, a strongly positive reward when the student passes the assessment of the target concept; 2) $r_{c-}$, a negative reward when the student fails the assessment of the target concept; 3) $r_{e+}$, a slightly positive reward when the student successfully comprehends an exercise; 4) $r_{e-}$, a slightly negative reward when the student fails to comprehend an exercise (too difficult) or has already mastered it (too easy); and 5) $r_\text{quit}$, a strongly negative reward when the student quits the online learning, either reaching the maximum turn $T$ or exceeding their patience threshold $\beta$.
\subsubsection{\textbf{User Simulation via Cognitive Diagnosis}}\label{sec:user-sim}
As the interactive tutoring is a dynamic process and it is costly and time-consuming to learn from human feedback, we follow previous policy learning studies in other interactive systems, such as interactive recommendation~\cite{sigir21-unicorn,recsys23-mm-crs} and task-oriented dialogues~\cite{convai2,sigdial22-tod-us}, to adopt an user simulator for training and evaluation. Existing studies \cite{kdd19-ktsim,ktsim-learningpath,ktsim-exercise} typically adopt knowledge tracing models \cite{ktsurvey21,www23-kt,www23-kt-lq} to simulate students' responses to the exercise. However, GITS further requires to assess their mastery of specific concepts. 

Cognitive diagnosis \cite{cd-survey,ijcai21-cd} is a fundamental technique in intelligent education, which aims to discover the proficiency level of students on specific concepts through the student performance prediction process. 
We adopt a widely-adopted cognitive diagnosis model, namely NeuralCD~\cite{aaai20-cdm}, as the simulator to predict the student's responses. 
After being trained on the student-exercise matrix $O$ and the concept-exercise matrix $Q$, NeuralCD can predict the probability $\rho$ within $[0,1]$ of a student $u$ correctly responding to an exercise $e$: 
\begin{equation}
    \rho_{u,e} = \mathbf{NeuralCD}(u,e),
\end{equation}
where the student $u$ is represented by the past interactions between students and exercises, while the exercise $e$ is represented by the association between each exercise and concept.

Then we estimate the student's mastery level of a concept by conduct an examination about the exercises related to the concept:
\begin{equation}
    d_{u,c} = (\sum_{e\in\mathcal{E}_{c}}\rho_{u,e})/|\mathcal{E}_c|
    \label{eq:master_score}
\end{equation}

There are three roles for the user simulator: 1) to determine the difficulty of an exercise to a student at a certain state; 2) to assess the student's mastery level of the concerned concept based on the student performance on the related exercises; and 3) to reflect the student's patience loss. 
Accordingly, given the predicted action $a_t$ at the current turn, we simulate the student response $f_t$ as well as obtain the reward $r_t$ and the patience loss $l_t$ as follows:
\begin{equation}
    \begin{cases}
    r_t=r_{c+},~ l_t =0,   & \quad \text{if } a_t = c^*, ~d_{u,c}\geq \delta \\
     r_t=r_{c-},~ l_t=1,  & \quad \text{if } a_t = c^*, ~d_{u,c}< \delta \\
    r_t=r_{e+},~ f_t=1,~ l_t=1-\rho_{u,a_t},     & \quad \text{if } a_t \in \mathcal{E}, ~ \lambda_- <\rho_{u,a_t} < \lambda_+ \\
    r_t=r_{e-},~ f_t=1,~ l_t=1-\rho_{u,a_t},     & \quad \text{if } a_t \in \mathcal{E},~ \rho_{u,a_t} \geq \lambda_+ \\
    r_t=r_{e-},~ f_t=-1,~ l_t=1-\rho_{u,a_t},     & \quad \text{if } a_t \in \mathcal{E},~ \rho_{u,a_t} \leq \lambda_- 
  \end{cases}
  \label{eq:reward}
\end{equation}
where $\delta$ denotes the threshold score of passing the examination regarding the target concept. $\lambda_+$ and $\lambda_-$ denote the interval of appropriate difficulty degree of the tutored exercise. 

We use the difficulty of the exercise, \textit{i.e.}, $1-\rho_{u,a_t} \in [0,1]$, to reflect the student $u$'s patience loss $l_t$ after being tutored with the exercise $e=a_t$. 
If the student fails the assessment, the largest patience loss is assigned to the student at this turn, \textit{i.e.}, $l_t=1$. 
The cumulative patience loss of the student at turn $t$ is calculated by $\mathcal{L}_t = \sum^t_{i=1} l_i$. 
If the cumulative patience loss exceeds the patience threshold ($\mathcal{L}_t\geq \beta$), the student will quit the online learning process. 
\subsubsection{\textbf{Dynamically Updating}}
In interactive ITS scenarios, the data is collected online, where students dynamically interact with various exercises, which can rarely meet the stationary condition in traditional cognitive diagnosis models \cite{kdd22-icd}. 
Therefore, we dynamically update the model parameters $\theta_\text{CD}$ of NeuralCD by applying gradient descent with the new exercise record at turn $t$:
\begin{equation}
    \theta_\text{CD} \leftarrow \theta_\text{CD} - \alpha \nabla y_t \log \rho_{u,a_t},
    \label{eq:update}
\end{equation}
where $\alpha$ denotes the learning rate of the dynamical update. Note that the binary label $y_t$ for incorrect responses in NeuralCD is set to 0, so we have $y_t=\max{(f_t,0)}$. In this manner, after successfully tutoring an exercise, the concept mastery degree of the simulated student will be improved accordingly. 

%In specific, we define the user's learning gain of comprehending an exercise $e_t$ at the timestep $t$ as $\mathcal{S}_c(u,c^*,s_t) - \mathcal{S}_c(u,c^*,s_{t-1})$, where $\mathcal{S}_c(\cdot)$ is a scoring function that evaluate the user's mastery degree of a certain concept. We define the user's patience loss of learning an exercise as contradictory to the user's knowledge level of the exercise, \textit{i.e.}, $1-\mathcal{S}_e(u,e_t,s_t)$, where $\mathcal{S}_e(\cdot)$ is a scoring function that evaluate the user's knowledge level of the exercise. 

\begin{algorithm}[t]
\caption{Training Procedure for \texttt{PAI}}
\label{algo}
\KwIn{The interaction data $\mathcal{D}$; the greedy probability $\epsilon$; the discounted factor $\gamma$; the maximum turn of conversations $T$; the patience threshold $\beta$; the learning rate $\alpha$;} 
\KwOut{The learned parameters $\theta_S$, $\theta_Q$; }
Initialize all parameters: $\{\mathbf{h}_i\}_{i\in \mathcal{N}}$;  $\theta_S$, $\theta_Q$\;%, $\theta_{Q'} \leftarrow \theta_Q$\;

\For{$\mathit{episode} = 1, 2, \ldots , N$}{

Get a sample ($u$, $c^*$) from $\mathcal{D}$;

Initialize state $\mathcal{G}_0=\mathcal{G}$, $\mathcal{H}_0=\varnothing$\;
Get candidate action space $\mathcal{A}_0$ via Action Selection\;
		
\For{turn $t = 0, 1, \ldots , T-1$}{
	
	Get state representation $f_{\theta_S}(s_t)$ via Eq.(\ref{state_rep})\;
	Select an action $a_t$ by $\epsilon$-greedy w.r.t Eq.(\ref{q-value})\;
	
	Receive reward $r_t$, feedback $f_t$, patience loss $l_t$ via Eq.(\ref{eq:reward})\;
        \If{$r_t=r_{c+}$ or $\mathcal{L}_t\geq \beta$}{break\;}
	Update the next state $s_{t+1} = \mathcal{T}(s_t,a_t,f_t)$ via Eq.(\ref{eq:transition})\;
        Update the user simulator via Eq.(\ref{eq:update})\;
	Get $\mathcal{A}_{t+1}$ via Action Selection\;
	Store $(s_t, a_t, r_t, s_{t+1}, \mathcal{A}_{t+1})$ to buffer $\mathcal{B}$\;
	Sample mini-batch of $(s_t, a_t, r_t, s_{t+1}, \mathcal{A}_{t+1})$ \;
	Compute the target value $y_t$ via Eq.~(\ref{target_value})\;
	Update $\theta_S$, $\theta_Q$ via SGD w.r.t the loss function Eq.(\ref{loss})\;
	%Update $\theta_{Q'}$ \;
}
}
\end{algorithm}

\subsection{Interaction}\label{sec:interaction}

\subsubsection{\textbf{Training}}
The tutoring policy is optimized by adopting the deep Q-learning network (DQN)~\cite{nature-dqn} to conduct reinforcement learning from interacting with the student. The training procedure of the \texttt{PAI} framework is presented in Algorithm~\ref{algo}.

During each episode in the GITS process, at each timestep $t$, the ITS agent obtains the current state representation $f_{\theta_S}(s_t)$ via the state representational learning described in Section~\ref{sec:state}. Then the agent selects an action $a_t$ with $\epsilon$-greedy from the candidate action space $\mathcal{A}_t$, which is obtained via the action selection strategies described in Section~\ref{sec:action_selection}. %Here we incorporate $\epsilon$-greedy method to balance the exploration and exploitation in action sampling (i.e., select a greedy action based on the max Q-value with probability $1-\epsilon$, and a random action with probability $\epsilon$).

%In specific, we employ the dueling Q-network~\cite{icml16-dueling} to compute the value function $f_{\theta_V}(\cdot)$ and advantage function $f_{\theta_A}(\cdot)$, respectively. Q-value $Q(s_t, a_t)$ is defined as the expected reward based on the state $s_t$ and the action $a_t$:  
In specific, we employ the dueling Q-network~\cite{icml16-dueling} to compute the Q-value $Q(s_t, a_t)$, which is defined as the expected reward based on the state $s_t$ and the action $a_t$:
\begin{equation}\label{q-value}
    Q(s_t, a_t) = f_{\theta_Q}(f_{\theta_S}(s_t), a_t),
\end{equation}
where $\theta_Q$ denotes the parameters in the dueling Q-network. 
%where $f_{\theta_V}(\cdot)$ and $f_{\theta_A}(\cdot)$ are two separate multi-layer perceptions with parameters $\theta_V$ and $\theta_A$, respectively, and let $\theta_Q = \{\theta_V, \theta_A\}$.

Then, the agent will receive the reward $r_t$ based on the user's feedback. According to the feedback, the current state $s_t$ transitions to the next state $s_{t+1}$, and the candidate action space $\mathcal{A}_{t+1}$ is updated accordingly. The experience $(s_t, a_t, r_t, s_{t+1}, \mathcal{A}_{t+1})$ is then stored into the replay buffer $\mathcal{B}$. To train DQN, we sample mini-batch of experiences from $\mathcal{B}$ via prioritized experience replay~\cite{iclr16-per}, and minimize the following loss function:
\begin{align}
    \mathcal{L}(\theta_Q, \theta_S) &= \mathbb{E}_{(s_t,a_t,r_t,s_{t+1},\mathcal{A}_{t+1}){\sim}\mathcal{B}}\bigl[(y_t {-} Q(s_t,a_t;\theta_Q,\theta_S))^2\bigr], \label{loss}\\
    y_t &= r_t + \gamma \max_{a_{t+1}\in\mathcal{A}_{t+1}} Q(s_{t+1},a_{t+1};\theta_Q,\theta_S), %\label{eq:yt}
    \label{target_value}
\end{align}
where $y_t$ is the target value based on the currently optimal $Q^*$ and $\gamma$ is a discounted factor for delayed rewards. In addition, we further adopt Double Q-learning~\cite{aaai16-doubledqn} to alleviate the overestimation bias problem in conventional DQN by employing a target network $Q'$ as a periodic copy from the online network. 

\iffalse
To alleviate the overestimation bias problem in conventional DQN, we adopt Double Q-learning~\cite{aaai16-doubledqn}, which employs a target network $Q'$ as a periodic copy from the online network. The target value of the online network is then changed to:
\begin{equation}
    y_t = r_t + \gamma Q'\big(s_{t+1}, \argmax_{a_{t+1}\in\mathcal{A}_{t+1}} Q(s_{t+1},a_{t+1};\theta_Q,\theta_S);\theta_{Q'},\theta_S\big),
\label{target_value}
\end{equation}
where $\theta_{Q'}$ denotes the parameter of the target network, which is updated by the soft assignment as:
\begin{equation}
    \theta_{Q'} = \tau\theta_{Q} + (1-\tau)\theta_{Q'},
\label{double}
\end{equation}
where $\tau$ is the update frequency.
\fi

\subsubsection{\textbf{Inference}} 
After training the \texttt{PAI} framework, given a student and his/her interaction history, we follow the same process to obtain the candidate action space and the current state representation, and then decide the next action according to max Q-value in Eq.(\ref{q-value}). If the selected action points to an exercise, the system will tutor the student on the exercise. Otherwise, the system will assess the student regarding the mastery degree of the target concept. 

\section{Experiment}
We conduct the experiments with respect to the following research questions (RQs):
\begin{itemize}
    \item \textbf{RQ1.} How is the overall performance of \texttt{PAI} comparing with heuristic planning, offline adpative learning, and vanilla RL-based baselines?
    \item \textbf{RQ2.} How does the cognitive structure learning affect the tutoring policy, including the graph-based state representation learning and prerequisite-guided action selection strategy?
    \item \textbf{RQ3.} How do different types of students affect the performance, \textit{e.g.}, different patience or different learning rates?
\end{itemize}

\begin{table}
\centering
  \caption{Summary statistics of datasets.}
%\begin{adjustbox}{max width=\linewidth}
%\setlength{\tabcolsep}{2mm}{
\begin{tabular}{lrrr}
\toprule 
& Computer Science & Math & Psychology  \\
\midrule 
 \#Users &1,966&2,701&4,710 \\
 \#Concepts &215&129&298  \\
 \#Exercises &1,631&1,019&1,177  \\
 \#Interactions &84,022&53,675&430,759 \\
 Avg. Exer./Con. &76&70&67 \\
\midrule 
\#Train Samples &24,946&33,248&295,028 \\
\#Test Samples &706&1,717&3,509 \\
\bottomrule
\label{dataset}
\end{tabular}%}
%\end{adjustbox}
\end{table}

\begin{table*}[t]
    \centering
    \caption{Experimental results. Success Rate: higher$\uparrow$ the better. Average Turn and Impatience: lower$\downarrow$ the better. $^\dagger$ indicates that the model is better than the best performance of baseline methods (\underline{underline} scores) with statistical significance (measured by paired significance test at $p<0.05$).}
    \begin{adjustbox}{max width=\textwidth}
    \begin{tabular}{lccccccccc}
    \toprule
         & \multicolumn{3}{c}{Computer Science} & \multicolumn{3}{c}{Math} & \multicolumn{3}{c}{Psychology} \\
         \cmidrule(lr){2-4}\cmidrule(lr){5-7}\cmidrule(lr){8-10}
         Method &  SR$\uparrow$ & AT$\downarrow$ & PL$\downarrow$&  SR$\uparrow$ & AT$\downarrow$ & PL$\downarrow$&  SR$\uparrow$ & AT$\downarrow$ & PL$\downarrow$\\
         \midrule
         %\multicolumn{7}{l}{\textit{Unsupervised Methods}}\\
         KNN&0.276&16.757&4.118&0.469&13.810&3.711&0.192&17.366&3.913 \\
         Greedy&0.273&16.508&4.052&\underline{0.491}&\underline{\textbf{13.519}}&3.696&0.196&17.309&3.961 \\
         \midrule
         \multicolumn{7}{l}{\textit{Supervised Methods}}\\
         DKT &	0.170&	17.830&	4.083&	0.213&	17.157&	4.120 &0.117 & 17.137 &4.133\\
         EKT	&0.267	&16.572&	3.689&	0.400&	14.636&	3.669& 0.173 & \underline{16.938} & 3.722\\
         \midrule
         \multicolumn{7}{l}{\textit{RL-based Methods}}\\
         DQN&	0.151&	17.207&	2.882&	0.366&	17.034&	\underline{2.063} &0.112 &17.124 &2.734\\
        %DDQN&	0.183&	14.014&	2.816&	0.479&	16.077&	2.257& \\
        %PDDQN&	0.221&	13.708&	3.009&	0.505&	15.551&	2.610& \\
         PDDDQN &\underline{0.314}&\underline{16.443}&\underline{\textbf{2.397}}&0.488&14.083&2.323&\underline{0.281}&16.941&\underline{\textbf{2.635}}\\
         \midrule
         \texttt{PAI}&\textbf{0.375}$^\dagger$&\textbf{16.223}$^\dagger$&2.851&\textbf{0.534}$^\dagger$&14.557&\textbf{2.131}$^\dagger$&\textbf{0.303}$^\dagger$&\textbf{16.903}&2.859\\
         \bottomrule
    \end{tabular}
    \end{adjustbox}
    \label{tab:overall}
\end{table*}

\subsection{Experimental Setups}
\subsubsection{\textbf{Datasets}}\label{sec:dataset}
We conduct experiments on three datasets in different subjects, including Computer Science, Math, and Psychology, extracted from the MOOCCubeX dataset\footnote{\url{https://github.com/THU-KEG/MOOCCubeX}} 
\cite{mooccubex}. 
In specific, each dataset includes the student-exercise records, exercise-concept relations, and concept prerequisite mappings. 
Following the common setting of recommendation evaluation \cite{uai09-bpr,www17-ncf}, we prune the users that have less than 15 records to reduce the data sparsity for the test set\footnote{Sparse user interaction data in test sets can significantly skew evaluation results by inflating or underrepresenting model performance. To mitigate this, preprocessing (e.g., pruning sparse users, as done in the original study) improves metric stability at the cost of reduced coverage, while the sparse user interaction data is worth studying for cold-start challenges.}. 
All the remaining records are adopted as the data for the offline pre-training of the node embeddings and the user simulator. 

After training the user simulator, we estimate the a student $u$'s mastery level of a concept $c$ based on the predicted performance on the exercises related to the concept via Eq.(\ref{eq:master_score}). According to the estimated mastery level, we divided the concepts into three difficulty level, including hard ($0.5<d_{u,c}<0.6$), medium ($0.6<d_{u,c}<0.7$), and easy ($0.7<d_{u,c}<0.8$). During the training and inference phases of RL, each sample is assigned with a target concept to start the interaction between the ITS and the student.  The statistics of three datasets are summarized in Table \ref{dataset}, where Computer Science and Math are relatively smaller datasets than Psychology. 
%Code and data will be released via \url{https://github.com/anonymous}.

%We randomly select a number of the concept set (0.6 < master level < 0.7) as the target concept for training, and the difficulty of it in the test set is marked as medium, the concept set(0.7<master level<0.8) is marked as easy, the concept (0.5<master level<0.6) is marked as hard. During training and testing, each user corresponds to a target concept

\subsubsection{\textbf{Baselines}}
As the GITS scenario is a new task, there are few suitable baselines. We compare our overall performance with two unsupervised planning baselines (KNN and Greedy), two offline supervised learning baselines (DKT and EKT), and two RL-based methods (DQN and PDDDQN):
\begin{itemize}
    \item \textbf{KNN} \cite{infsci18-learningpath}. The agent selects the action based on the nearest cosine distance of the candidate action and the user embeddings. \citet{infsci18-learningpath} use KNN to conduct the collaborative filtering for learning path recommendation. 
    \item \textbf{Greedy}. The agent randomly selects the exercise that is related to the target concept to tutor. After the user comprehends an exercise, the agent will assess the mastery of the target concept.  
    %\item \textbf{DKT-Greedy}. The agent selects the exercise from the Greedy candidate to tutor based on the user's knowledge level predicted by the DKT model. 
    %\item \textbf{Random-SPP}. The agent randomly selects the exercise from the candidates to tutor. The candidate exercises are those related to the concepts in the shortest prerequisite path (SPP) towards the target concept. 
    %\item \textbf{DKT-SPP}. The agent selects the exercise from the SPP candidates to tutor based on the user's knowledge level predicted by the DKT model. 
    \item \textbf{DKT} \cite{dkt} and \textbf{EKT} \cite{ekt}. The agent selects the exercise to tutor based on the user's knowledge level predicted by a knowledge tracing model trained on the student-exercise interaction history data, including DKT and EKT. After the user comprehends an exercise, the agent will assess the mastery of the target concept.  
    \item \textbf{DQN} \cite{nature-dqn}. The agent selects the action based on the Q-value computed by DQN trained on the same RL process as \texttt{PAI}. 
    We further compare to an advanced DQN, incorporating double DQN \cite{aaai16-doubledqn}, dueling network \cite{icml16-dueling}, and prioritized experience replay \cite{iclr16-per}, namely \textbf{Prioritized Dueling Double DQN (PDDDQN)}.
    %We further compare to several advanced DQNs, including  \textbf{Double DQN (DDQN)} \cite{aaai16-doubledqn},  \textbf{Prioritized DDQN (PDDQN)} \cite{iclr16-per}, and \textbf{Prioritized Dueling DDQN (PDDDQN)} \cite{icml16-dueling}. 
\end{itemize}

\subsubsection{\textbf{Evaluation Metrics}}
For evaluation protocols, we adopt the Success Rate (SR) to measure the ratio of successfully reaching the educational goal by turn $T$ for measuring the effectiveness of the GITS. Besides, the Average Turns (AT) is used to measure the efficiency of reaching the goal. In addition, in online education, the student's learning interest is also an important criteria for a successful online learning session. To this end, we adopt the student's patience loss (PL), which is determined by the cumulative difficulty of exercises recommended to the student, for evaluation. 

\subsubsection{\textbf{Implementation Details}}
We adopt TransE \cite{transe} from OpenKE \cite{openke} to pretrain the node embeddings in the constructed cognitive graph with all the interaction records. 
For evaluation, we set the maximum turn $T$ as 20 and the patience threshold $\beta$ as 4. The threshold score $\delta$ of passing the examination is set to 0.9. The interval $[\lambda_-,\lambda_+]$ of appropriate difficulty degree of the tutored exercise is set to [0.5,1]. 
We set the rewards as follows: $r_{c+}=1$, $r_{c-}=-0.1$, $r_{e+}=0.01$, $r_{e-}=-0.1$, and $r_\text{quit}=-0.3$. 
The strong positive reward for goal completion ($r_{c+}=1$) and the small negative reward ($r_{c-}=r_{e-}=-0.1$) for encouraging efficiency follow the typical setting in the RL-based methods of other task-oriented interactive systems, such as task-oriented dialogues \cite{dpl-survey} and conversational recommendation \cite{ear}, while the other two are tuned in a small validation set. 

The hyperparameters have been empirically configured as follows: The embedding size and the hidden size are respectively set to be 64 and 100. 
The number of GCN layers $L$ is fixed at 2. The number of candidate exercises $N$ is set to 30. The learning rate $\alpha$ for dynamically updating the user simulator is set to 0.02. During the training procedure of DQN, the experience replay buffer has a capacity of 50,000, and the mini-batch size is set to 128. The learning rate and $L_2$ norm regularization are adjusted to 1e-4 and 1e-6, respectively. The discount factor, $\gamma$, is assigned values of 0.999.

\subsection{Experimental Results (RQ1)}

\subsubsection{\textbf{Overall Evaluation}}

Table \ref{tab:overall} summarizes the experimental results of the proposed method, \texttt{PAI}, and all baselines across three datasets. 
In general, \texttt{PAI} outperforms the baselines by achieving a higher success rate and less average turns except for the average turn in the Math dataset, indicating the effectiveness and efficiency of \texttt{PAI} in tackling the GITS task.  
As for the baselines, we observe a common drawback that they all may struggle to determine when to assess the student's mastery level of the target concept. 
For example, the Impatience scores for KNN and Greedy consistently reach or exceed the patience threshold (\textit{i.e.}, $\beta=4$). This indicates that a significant number of students discontinue the online learning process due to the repeated assessment of the target concept at an inappropriate time, depleting their patience entirely. 
In contrast, PDDDQN exhibits the lowest Impatience score but fails to compete with \texttt{PAI} in terms of Success Rate and Average Turn, since PDDDQN fails to take the action of assessment even when the student has already mastered the target concept. 
As for the Psychology dataset, given more training samples for RL-based methods, the performance gap from heuristic planning baselines becomes much more significant than the other two datasets. 
%Due to the space limitation, we place the qualitative analysis of the method performance in Appendix \ref{app:qualitative}, including expert ratings and case studies. 

\input{fig/turn}

\subsubsection{\textbf{Performance w.r.t Different Turns}}
Besides the final success rate at the maximum number of interaction turns (\textit{i.e.}, $T=20$), we also present the performance comparison of success rate at each turn in Figure \ref{fig:turn}. 
In order to better observe the differences among different methods, we report the relative success rate compared with the baseline PDDDQN. For example, the line of $y=0$ represents the curve of success rate for PDDDQN against itself. 
The results show that Greedy and KNN substantially outperform two RL-based methods (\textit{incl.}, PDDDQN and \texttt{PAI}) at the early phases. 
They may indeed yield positive results by effectively addressing simpler, easy-to-grasp concepts during the initial phases of the interactive learning process. This can result in relatively strong performance. However, as the complexity of the tasks increases, their performance diminishes rapidly. 
Overall, the proposed \texttt{PAI} proves to be a highly effective approach. It surpasses all baseline methods, particularly in the later stages of the interactive learning process. \texttt{PAI} accomplishes this by guiding the student through a personalized sequence of educational interactions, ultimately enabling them to master the target concept. 
Moreover, the results also shed light on a promising future direction to investigate a hybrid policy that can adaptively adjust the tutoring policy according to the difficulty of the target concept.

\subsubsection{\textbf{Evaluation on User Simulation}}

Our user simulation relies on an existing cognitive diagnosis model, NeuralCD \cite{aaai20-cdm}. Despite the effectiveness of this model validated in other datasets, it is necessary to evaluate the reliability of this prediction in our studied scenarios. To this end, during the training phase of the user simulator, we leave out a testing set for validation. Following the original study \cite{aaai20-cdm}, we adopt evaluation metrics from both classification aspect and regression aspect, including Accuracy, RMSE (root mean square error), and AUC (area under the curve). The evaluation results are presented in Table \ref{tab:neuralcd}, which indicates a promising performance on the datasets across three subjects. Compared to the original datasets for cognitive diagnosis, the MOOCCubeX dataset is more concentrated on specific subjects and interrelated concepts. Therefore, it could be much easier for a powerful cognitive diagnosis model to handle such scenarios.

\begin{table}[]
    \centering
    \caption{Evaluation results on user simulation. }
    \begin{tabular}{lccccccccc}
    \toprule
         & \multicolumn{3}{c}{Computer Science} & \multicolumn{3}{c}{Math} & \multicolumn{3}{c}{Psychology} \\
         \cmidrule(lr){2-4}\cmidrule(lr){5-7}\cmidrule(lr){8-10}
        & Acc$\uparrow$ & RMSE$\downarrow$ & AUC$\uparrow$& Acc$\uparrow$ & RMSE$\downarrow$ & AUC$\uparrow$& Acc$\uparrow$ & RMSE$\downarrow$ & AUC$\uparrow$\\
         \midrule
         NeuralCD& 0.890 & 0.294 & 0.898 & 0.854 & 0.323& 0.906 & 0.867& 0.309 & 0.910\\
         \bottomrule
    \end{tabular}
    %\end{adjustbox}}
    \label{tab:neuralcd}
\end{table}

\begin{table}[]
    \centering
    \caption{Ablation study. }
    \begin{tabular}{lcccccc}
    \toprule
         & \multicolumn{3}{c}{Computer Science} & \multicolumn{3}{c}{Math}  \\
         \cmidrule(lr){2-4}\cmidrule(lr){5-7}
         Method  & SR$\uparrow$ & AT$\downarrow$ & PL$\downarrow$& SR$\uparrow$ & AT$\downarrow$ & PL$\downarrow$\\
         \midrule
         \texttt{PAI}&\textbf{0.375}&16.223&2.851&\textbf{0.534}&14.557&\textbf{2.131}\\
         \midrule
         - w/o Cognitive Graph &0.365&\textbf{16.037}&2.797&0.529&\textbf{13.784}&2.318\\
         - w/o Graph Pretrain&0.212&17.948&3.119&0.398&16.010&2.276 \\
         \midrule
         - w/o  Action Selection &0.257&17.420&2.719&0.330&16.155&2.285\\
        - w/o Prerequisite&0.332&16.867&\textbf{2.612}&0.491&15.211&2.132 \\
         \bottomrule
    \end{tabular}
    %\end{adjustbox}}
    \label{tab:ablation}
\end{table}

\subsection{Ablation Study (RQ2)}

\subsubsection{\textbf{Graph-based State Representation Learning}}
As for the cognitive graph for state representation learning, we investigate two variants: 1) discarding the graph-based pre-trained knowledge by using randomly initialized node embeddings (-w/o Graph Pretraining); and 2) discarding the cognitive graph structure by using the original node embeddings (-w/o Cognitive Graph). 
As presented in Table \ref{tab:ablation}, the results clearly show that relying on online training to build node representations from scratch in the cognitive graph is challenging. This is evident from the substantial performance drop observed when using randomly initialized node embeddings. 
However, the enhancement gained from using GCN to refine node representations in the cognitive graph is marginal. 
To sum up, the performance of \texttt{PAI} actually benefits from the cognitive graph structure, but the results also suggest the need for further exploration of more effective graph learning approaches for this problem.

\begin{table*}[]
    \centering
    \caption{Performance in terms of concepts with different difficulty levels (Diff.). The \textbf{bold} scores denote the best performance on the same difficulty level for each metric. The \colorbox{gray!30}{shadowed} scores denote the best performance with the same method for each metric.}
    \begin{adjustbox}{max width=\textwidth}
    \begin{tabular}{llccccccccc}
    \toprule
         && \multicolumn{3}{c}{Computer Science} & \multicolumn{3}{c}{Math} & \multicolumn{3}{c}{Psychology} \\
         \cmidrule(lr){3-5}\cmidrule(lr){6-8}\cmidrule(lr){9-11}
         Method & Diff. & SR$\uparrow$ & AT$\downarrow$ & PL$\downarrow$&  SR$\uparrow$ & AT$\downarrow$ & PL$\downarrow$&  SR$\uparrow$ & AT$\downarrow$ & PL$\downarrow$\\
         \midrule
         \rowcolor[gray]{0.85}\multirow{3}{*}{KNN} & Easy &0.409&14.949&3.974&0.577&12.293&3.538&0.353&\textbf{15.293}&3.656\\ 
         & Medium &0.243&16.954&4.134&0.472&13.804&3.714&0.183&17.363&3.885\\
         & Hard&0.073&18.982&4.295&0.282&16.362&3.997&0.012&19.834&4.263\\
         \midrule
         \rowcolor[gray]{0.85}\multirow{3}{*}{Greedy} & Easy &0.409&\textbf{14.732}&3.825&\textbf{0.651}&\textbf{11.268}&3.395&0.352&15.297&3.827\\
         & Medium&0.272&16.538&4.069&0.492&\textbf{13.590}&3.736&0.197&17.208&4.037\\
         & Hard&0.073&19.006&4.345&0.218&17.164&4.131&0.010&19.858&4.428\\
         \midrule
         \rowcolor[gray]{0.85}\multirow{3}{*}{PDDDQN} & Easy&0.404&15.270&\textbf{1.453}&0.541&13.366&1.738&0.337&16.100&\textbf{1.960} \\
         & Medium&0.302&16.574&\textbf{2.614}&0.482&14.478&2.226&0.333&16.469&\textbf{2.426}\\
         & Hard&0.206&\textbf{17.875}&\textbf{3.340}&0.411&\textbf{14.573}&3.463&0.132&18.693&\textbf{3.769}\\
         \midrule
         \midrule
         \rowcolor[gray]{0.85}\multirow{3}{*}{\texttt{PAI}} & Easy&\textbf{0.439}&15.413&2.211&0.570&14.545&\textbf{1.732}&\textbf{0.370}&15.986&1.973 \\
         & Medium&\textbf{0.413}&\textbf{15.892}&2.935&\textbf{0.534}&14.553&\textbf{1.971}&\textbf{0.351}&\textbf{16.467}&2.858\\
         & Hard&\textbf{0.213}&18.000&3.606&\textbf{0.472}&14.582&\textbf{3.073}&\textbf{0.146}&\textbf{18.690}&3.915\\
         \bottomrule
    \end{tabular}
    \end{adjustbox}
    \label{tab:difficulty}
\end{table*}

\subsubsection{\textbf{Prerequisite-guided Action Selection}}
As for the prerequisite relations for action selection, we also investigate two variants: 1) discarding the guidance of prerequisite-based concept scores by only using the exercise score for action selection (-w/o Prerequisite); and 2) discarding the action selection strategy by regarding the whole exercise set as the candidate action space (-w/o Action Selection). 
The results in Table \ref{tab:ablation} show that the performance of \texttt{PAI} suffers a noticeable decrease when the action selection is omitted. This underscores the substantial contribution of the proposed action selection strategy to the sampling efficiency in the RL framework. Specifically, the inclusion of prerequisite guidance further enhances performance by providing valuable prior knowledge.

\subsection{Adaptability Analysis (RQ3)}
Education is not one-size-fits-all, and learners possess diverse backgrounds and aptitudes. 
In order to gain insights into the adaptability and effectiveness of ITSs in catering to a wide range of students (from beginners to advanced learners, from apathetic learners to enthusiastic learners, from deliberate learners to fast learners, etc), we conduct several analysis by varying the simulation settings. 
%By assessing how an ITS handles concepts of different complexities and different types of students, we gain insights into its adaptability and effectiveness in catering to a wide range of students, from beginners to advanced learners, from apathetic learners to enthusiastic learners, from deliberate learners to fast learners, etc. We conduct several adaptability analysis of ITSs by varying the simulation settings in Section \ref{sec:user-sim}. 

\subsubsection{\textbf{Target Concepts with Different Difficulty Levels}}\label{sec:difficulty}
We first analyze the effect of concept difficulties by dividing the target concepts into three levels, including easy, medium, and hard, as introduced in Section \ref{sec:dataset}. 
The evaluation results are summarized in Table \ref{tab:difficulty}. 
There are several notable observations as follows. 
(1) Overall, the performance of all approaches drops when increasing the difficulty of the target concepts. 
(2) KNN and Greedy exhibit stronger capabilities in handling easy-to-learn concepts than RL-based methods (\textit{incl.}, PDDDQN and \texttt{PAI}), since the students can easily master the concepts by learning just few random exercises that is related to the target concept, which may downgrade the necessity of a tailored plan of learning path. 
(3) Conversely, KNN and Greedy merely work when handling hard-to-learn concepts, while RL-based methods perform much better, indicating the importance of content planning for more robust and capable ITSs.

\subsubsection{\textbf{Students with Different Patience Thresholds}}

\input{fig/patience}

In the main experiment, we set the patience threshold ($\beta$) of the simulated student as 4. In reality, there are both motivated students who has a high patience threshold and apathetic students who are easier to quit the learning with a low patience threshold. 
We analyze the effect of the learning patience by changing the patience threshold ($\beta$) of the simulated student within [1, 2, 3, 4, 5]. 
As expected, the results depicted in Figure \ref{fig:patience} indicate a general trend: the success rate tends to increase as the students' patience thresholds rise. In the case of students with low patience thresholds, \texttt{PAI} exhibits a significant performance advantage over KNN and Greedy, consistently outperforming PDDDQN. However, KNN and Greedy show a faster rate of increase in performance compared to PDDDQN and \texttt{PAI}. This leads to their superiority when students possess a high level of patience (e.g., $\beta>5$) in learning despite encountering setbacks.

\subsubsection{\textbf{Students with Different Learning Rates}}

\input{fig/learn_rate}

In the main results, we set the learning rate ($\alpha$) for the dynamic update of the user simulator at 0.02, which serves as a simulation of the student's learning speed. However, in real-life scenarios, students indeed exhibit varying learning speeds. To investigate the impact of the learning speed, we conducted an analysis by altering the learning rate ($\alpha$) of the simulated student, considering values within the range of [0.01, 0.02, 0.03, 0.04, 0.05]. 
As shown in Figure \ref{fig:learn_rate}, we observe a clear tendency that simulated students with higher learning rates are more likely to succeed in mastering the target concept when using KNN, PDDDQN, and \texttt{PAI}. This trend aligns with real-world scenarios. In contrast, the Greedy approach yields more inconsistent results. 
Specifically, \texttt{PAI} consistently surpasses PDDDQN in interacting with students varying learning rate. 
Compared with KNN, although KNN outperforms \texttt{PAI} for those students with high learning rate (\textit{e.g.}, $\alpha\geq0.03$), these students also suffer from losing more patience during their online learning experience than interacting with \texttt{PAI}.

\subsubsection{\textbf{Summary}}
In conclusion, the concept difficulty, the student's patience threshold and learning rate all play crucial roles in shaping the effectiveness of various learning strategies, highlighting the need for adaptive and proactive ITS approaches for addressing different learning challenges. 
This also underscores the significance of investigating the diversity in students' learning patterns, recognizing that education cannot follow a one-size-fits-all approach. The proposed datasets and the user simulator offer a valuable testbed for exploring this phenomenon. 

\subsection{Qualitative Analysis}\label{app:qualitative}
Apart from the automatic evaluation on effectiveness and efficiency of different methods, we further conduct qualitative analysis to investigate different aspects of the methods via human evaluations. 

\subsubsection{Expert Ratings}
\begin{figure}[]
   \centering  
   \includegraphics[width=0.6\textwidth]{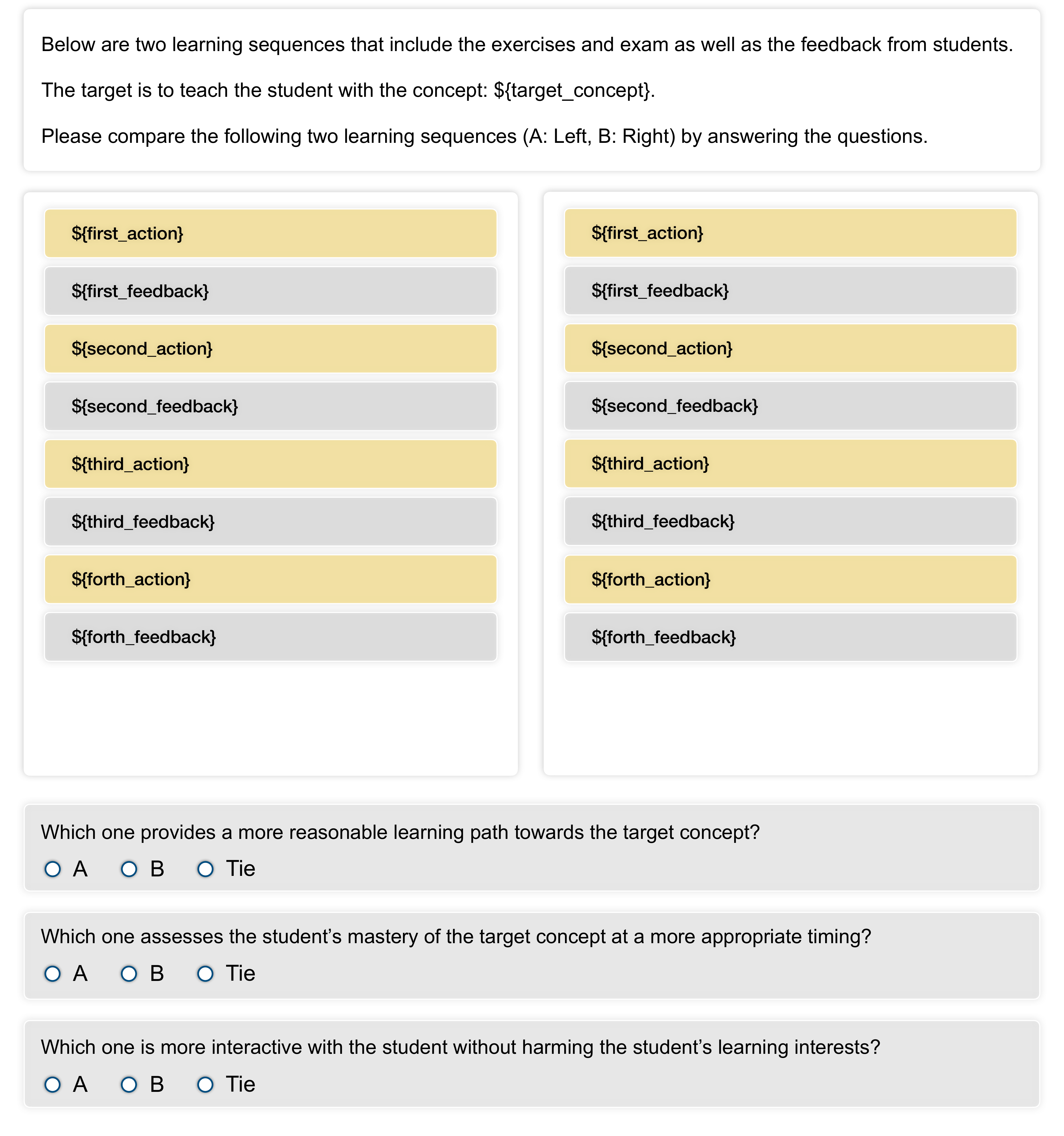} 
   
   \caption{User interface (UI) used for human evaluation.} 
   \label{fig:expert}
    %\vspace{0.2cm}
\end{figure}

\begin{sidewaysfigure*}[]
   \centering  
   \includegraphics[width=\textwidth]{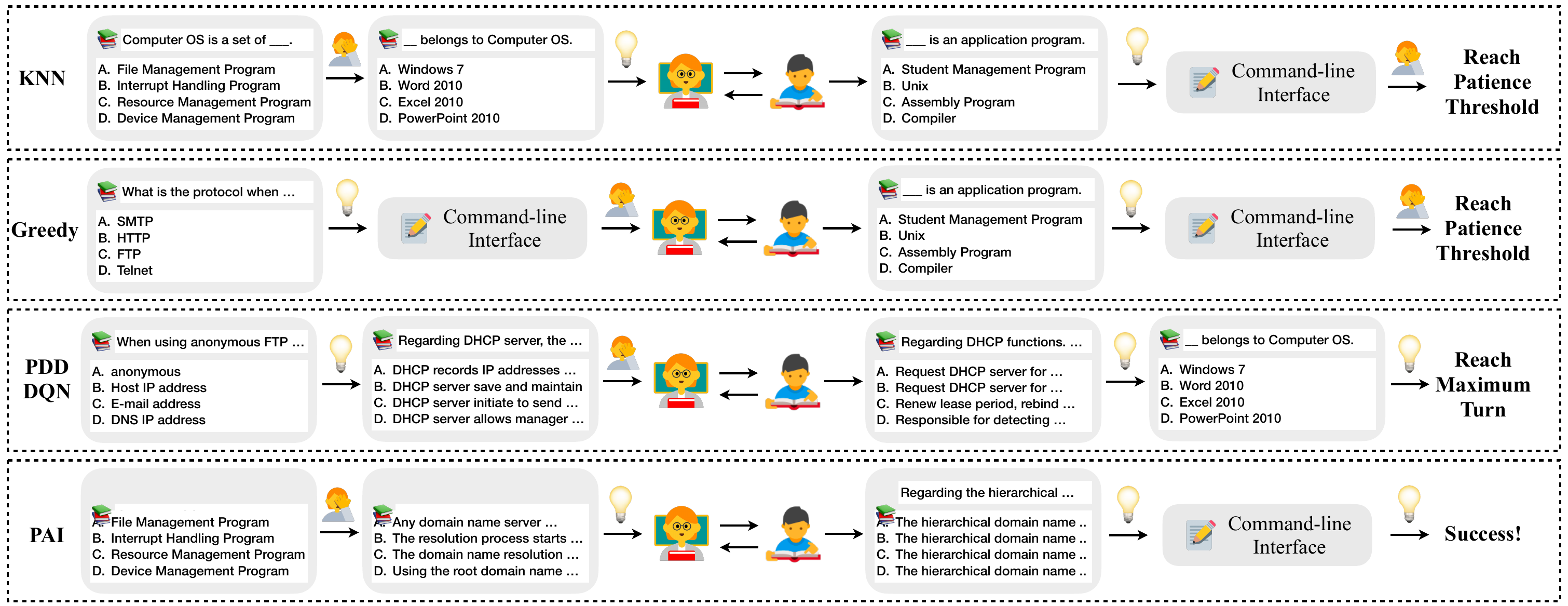} 
   
   \caption{Case study. The target concept is set to be "Command-line Interface".} 
   \label{fig:case}
    %\vspace{0.2cm}
\end{sidewaysfigure*}

Following previous studies in adaptive learning \cite{kbs-expert,kdd19-ktsim}, we invite two experts in specific subjects to evaluate the planning results of different methods based on their experiences. 
We randomly sample 50 cases for expert ratings. 
Experts are asked to compare learning sequences produced by \texttt{PAI} and another baseline by rating which one is better in terms of three perspectives with \textit{Win}/\textit{Tie}/\textit{Lose}. The user interface template used for human evaluation is presented in Figure \ref{fig:expert}. Two outputs are shown side by side, and the order is random. The example learning sequences are illustrated in Figure \ref{fig:case}.
Here we consider the following perspectives: 
\begin{itemize}
    \item Planning: Which one provides a more reasonable learning path towards the target concept?
    \item Assessment: Which one assesses the student's mastery of the target concept at a more appropriate timing?
    \item Interaction: Which one is more interactive with the student without harming the student's learning interests? 
\end{itemize}

\begin{table}[]
    \centering
    \caption{Expert ratings. The Fleiss’ kappa of the annotations is 0.71, which indicates ``substantial agreement”, and the final scores are calculated by average.}
    \begin{tabular}{lccccccccc}
    \toprule
        & \multicolumn{3}{c}{Planning} & \multicolumn{3}{c}{Assessment} & \multicolumn{3}{c}{Interaction} \\
        \cmidrule(lr){2-4}\cmidrule(lr){5-7}\cmidrule(lr){8-10}
        \texttt{PAI} vs. & Win & Tie & Lose & Win & Tie & Lose& Win & Tie & Lose\\
        \midrule
        KNN &  \textbf{45\%} & 31\% & 24\% & \textbf{51\%} &  26\% &23\% & \textbf{59\%} & 23\% & 18\% \\
        Greedy & \textbf{62\%} & 26\% & 12\% & \textbf{53\%} & 36\% &  11\% & \textbf{77\%} & 16\% &7\%\\
        PDDDQN & 32\% & \textbf{39\%} & 29\% & \textbf{46\%} & 26\% &  28\% & 37\%& \textbf{46\% }&17\% \\
    \bottomrule
    \end{tabular}%}
    \label{tab:expert}
    %\vspace{0.2cm}
\end{table}

Table \ref{tab:expert} presents a summary of expert ratings, offering valuable insights into the comparative performance of different methods.
It is evident that \texttt{PAI} consistently outperforms KNN and Greedy when evaluated from three distinct perspectives. Notably, in terms of the perspective of Interaction, \texttt{PAI} excels by providing a significantly more engaging learning experience, addressing the common issue of student impatience encountered when interacting with KNN and Greedy. 
PDDDQN, on the other hand, demonstrates competitive performance alongside \texttt{PAI}. However, it is worth noting that, based on the Assessment score, \texttt{PAI} excels in determining the optimal moment to conduct assessments related to a student's mastery of the target concept, contributing to its comprehensive advantage.

\subsubsection{Case Study}
In order to intuitively present the comparison among different methods, we illustrate a representative case in Figure \ref{fig:case}. Since the interactions are typically more than 10 turns, we only present part of the learning sequence. 
In this case, the target concept is "Command-line Interface". 
We observe that KNN and Greedy  often evaluate a student's grasp of the target concept prematurely, potentially affecting the student's motivation negatively. This impatience can lead to the student discontinuing their efforts to understand the concept. 
In contrast, PDDDQN operates in an entirely opposite manner by persistently assigning exercises, even when a student may have already achieved mastery. This tendency can lead to a surplus of tasks, reaching the maximum allowed number of interactions. 
In summary, the proposed \texttt{PAI} is designed to orchestrate a more effective sequence of actions to attain the target concept. This approach aims to strike a balance between assessment and engagement, optimizing the learning experience.

\section{Conclusion}
In this work, we emphasize the importance of proactive engagement in interactive ITSs to enhance online education. The introduction of GITS exhibits a new task that requires the ITS to proactively plan customized sequences of exercises and assessments, facilitating students' mastery of specific concepts. To address the challenge of goal-oriented policy learning in GITS, we introduced a graph-based reinforcement learning framework, named \texttt{PAI}. \texttt{PAI} utilizes cognitive structure information to improve state representation and action selection, taking into account both exercise tutoring and concept assessment. Additionally, we create three benchmark datasets spanning various subjects to support further academic research on GITS. Experimental results show the effectiveness and efficiency of \texttt{PAI}, with comprehensive adaptability analyses conducted to evaluate its performance across diverse students.

It is worth noting that our primary emphasis lies in strategizing proactive tutoring rather than delving into dialogue generation or other modalities. However, it's essential to acknowledge certain limitations. Our study does not encompass the specific interface designs and potential errors related to content understanding and generation.  Additionally, owing to the high expenses for real user studies, our evaluation primarily revolves around user simulations. To ensure a broad range of simulation scenarios, we have undertaken a comprehensive analysis across various types of simulated students as well as qualitative analysis with human evaluation.

\section*{Acknowledgment}
This research was supported by the Singapore Ministry of Education (MOE) Academic Research Fund (AcRF) Tier 1 grant (No. MSS24C004, No. MSS24C012).

\balance
\bibliographystyle{ACM-Reference-Format}
\bibliography{sample-base}
\end{document}

%% file: fig/turn.tex
\begin{figure}[t]
    \centering
    \begin{tikzpicture}
    \pgfplotsset{footnotesize,samples=10}
    \begin{groupplot}[group style = {group size = 2 by 2, horizontal sep = 20pt, vertical sep= 35pt}, width = 6.4cm, height = 4.0cm]

\nextgroupplot[scaled ticks=false, tick label style={/pgf/number format/fixed}, title = {Computer Science},
legend style = { column sep = 10pt, legend columns = 1, legend to name = grouplegend}, 
    ylabel={Relative Success Rate}, 
            xlabel={Turn of Interaction},
    xmin=0, xmax=21,
    ymin=-10, ymax=10,
    xtick={0,5,10,15,20},
    ytick={-10,-5,0,5,10}, grid=both,
    grid style={dashed, gray!50},]

\addplot[
    color=brown, mark=diamond, mark options={scale=1}
    ]
    coordinates {
    (1, 0.0)(2, 0.0)(3, 0.0)(4, 0.22192003383910394)(5, 0.9619614275753916)(6, 0.8758702241062053)(7, 2.553075716419792)(8, 4.7741832435993805)(9, 5.628822330809255)(10, 4.477244910590167)(11, 3.4736735361797026)(12, 2.6087235556415456)(13, 1.3139145597209523)(14, 0.8803603218416334)(15, 0.1603429325592931)(16, -0.5592169770568756)(17, -1.133386815751053)(18, -2.4281958116716575)(19, -2.8580548270540964)(20, -3.7230048075922504)
    };

\addplot[
    color=purple, mark=x, mark options={scale=1}
    ]
    coordinates {
    (1, 0.0)(2, 0.0)(3, 0.0)(4, -0.4298590153824378)(5, -0.43809935900478114)(6, -2.3116353886151484)(7, -2.5298701347820476)(8, -5.187353314008064)(9, -0.8308658022496362)(10, 4.122877549926901)(11, 3.119306175516437)(12, 2.2543561949782798)(13, 0.9595471990576865)(14, 0.5259929611783676)(15, -0.1940244281039727)(16, -0.9135843377201414)(17, -1.4877541764143187)(18, -2.782563172334923)(19, -3.2124221877173618)(20, -4.077372168255517)
    };

\addplot[
    color=red,
    mark=o,mark options={scale=0.5}
    ]
    coordinates {
    (1, 0.0)(2, 0.0)(3, 0.0)(4, 0.0)(5, 0.0)(6, 0.0)(7, 0.0)(8, 0.0)(9, 0.0)(10, 0.0)(11, 0.0)(12, 0.0)(13, 0.0)(14, 0.0)(15, 0.0)(16, 0.0)(17, 0.0)(18, 0.0)(19, 0.0)(20, 0.0)
    };

\addplot[
    color=blue,
    mark=square,mark options={scale=0.5}
    ]
    coordinates {
    (1, 0.0)(2, 0.0)(3, 0.0)(4, -0.4298590153824378)(5, -0.8612547580412793)(6, -1.7255829706353663)(7, -3.7368784216193283)(8, -6.3431072221126925)(9, -5.036297599369124)(10, -3.1725507145276275)(11, -0.013373065821470353)(12, 1.9946846423317788)(13, 2.7147020316141193)(14, 4.291363997108533)(15, 5.006313724895489)(16, 5.577410109036868)(17, 5.725416154845508)(18, 5.581105331533742)(19, 6.300665241149922)(20, 6.1541959226177)
    };

    \addlegendentry{KNN}
    \addlegendentry{Greedy}
    \addlegendentry{PDDDQN}
    \addlegendentry{\texttt{PAI}}
    
\nextgroupplot[scaled ticks=false, tick label style={/pgf/number format/fixed}, title = {Math},
legend style = { column sep = 10pt, legend columns = -1, legend to name = grouplegend2}, 
            xlabel={Turn of Interaction},
    xmin=0, xmax=21,
    ymin=-15, ymax=20,
    xtick={0,5,10,15,20},
    ytick={-15,-10,0,10,20}, grid=both,
    grid style={dashed, gray!50},]

\addplot[
    color=brown, mark=diamond,mark options={scale=1}
    ]
    coordinates {
    (1, 0.0)(2, 0.0)(3, 0.0)(4, 1.209456348914636)(5, 0.506303438881478)(6, 3.52538724984152)(7, 10.482533125959026)(8, 16.659522335478822)(9, 11.950113953144626)(10, 9.677642866604124)(11, 7.760157032398174)(12, 5.33883428432651)(13, 3.673435918321716)(14, 2.260692210876214)(15, 1.7061776978929244)(16, 0.4959338389818102)(17, -0.10931556507443041)(18, -0.7642800633981373)(19, -1.3187372844260636)(20, -1.9238019219262437)
    };

\addplot[
    color=purple, mark=x,mark options={scale=1}
    ]
    coordinates {
    (1, 0.0)(2, 0.0)(3, 0.0)(4, -0.656424683633063)(5, -3.230509131264425)(6, -4.443187385399784)(7, -4.097713787621221)(8, 4.529331538621367)(9, 10.911954581651912)(10, 11.865072720700631)(11, 9.94758688649468)(12, 7.5262641384230164)(13, 5.860865772418222)(14, 4.44812206497272)(15, 3.893607551989431)(16, 2.6833636930783165)(17, 2.078114289022076)(18, 1.4231497906983692)(19, 0.8686925696704428)(20, 0.2636279321702628)
    };

\addplot[
    color=red,
    mark=o,mark options={scale=0.5}
    ]
    coordinates {
    (1, 0.0)(2, 0.0)(3, 0.0)(4, 0.0)(5, 0.0)(6, 0.0)(7, 0.0)(8, 0.0)(9, 0.0)(10, 0.0)(11, 0.0)(12, 0.0)(13, 0.0)(14, 0.0)(15, 0.0)(16, 0.0)(17, 0.0)(18, 0.0)(19, 0.0)(20, 0.0)
    };

\addplot[
    color=blue,
    mark=square,mark options={scale=0.5}
    ]
    coordinates {
    (1, 0.0)(2, 0.0)(3, 0.0)(4, -0.5053923904538427)(5, -4.3885752393049176)(6, -7.0675699889268975)(7, -9.439694702628252)(8, -10.190003040770529)(9, -10.548915656645312)(10, -7.916384682823862)(11, -5.6915819354454165)(12, -4.376718669142249)(13, -2.5616966325077706)(14, -0.9994123277678457)(15, 0.7660091985098871)(16, 1.1186741266521694)(17, 2.1265585023195364)(18, 3.2359699520881744)(19, 4.4985614706157415)(20, 4.5491688870880065)
    };
    
\nextgroupplot[scaled ticks=false, tick label style={/pgf/number format/fixed}, title = {Psychology},
legend style = { column sep = 10pt, legend columns = -1, legend to name = grouplegend2}, 
            xlabel={Turn of Interaction},
            ylabel={Relative Success Rate},
    xmin=0, xmax=21,
    ymin=-10, ymax=5,
    xtick={0,5,10,15,20},
    ytick={-10,-5,0,5}, grid=both,
    grid style={dashed, gray!50},]

\addplot[
    color=brown, mark=diamond,mark options={scale=1}
    ]
    coordinates {
    (1, 0.0)(2, 0.0)(3, 0.3310613437195716)(4, 1.4214416057639214)(5, 1.4377021453377186)(6, 0.32333792255330107)(7, 2.4340765711335517)(8, 2.795296892139713)(9, 4.633776300939552)(10, 2.0815292771511813)(11, -0.9174995182987145)(12, -3.0984345445641273)(13, -4.441875255614872)(14, -5.142869328720234)(15, -5.843996197340448)(16, -6.681707935942146)(17, -7.3050652186288065)(18, -7.889532983571243)(19, -8.396096109207043)(20, -8.922124481687957)
    };

\addplot[
    color=purple, mark=x,mark options={scale=1}
    ]
    coordinates {
    (1, 0.0)(2, 0.0)(3, 0.2928168740279938)(4, 1.4644258282826144)(5, 1.9462347981940187)(6, 0.8136312765943503)(7, 2.223749113061191)(8, 3.029441841495041)(9, 4.207514941902884)(10, 2.529722356881925)(11, -0.4694344783358201)(12, -2.6503695046012328)(13, -3.993810215651977)(14, -4.69480428875734)(15, -5.395931157377554)(16, -6.233642895979252)(17, -6.857000178665912)(18, -7.441467943608348)(19, -7.948031069244149)(20, -8.474059441725062)
    };

\addplot[
    color=red,
    mark=o,mark options={scale=0.5}
    ]
    coordinates {
    (1, 0.0)(2, 0.0)(3, 0.0)(4, 0.0)(5, 0.0)(6, 0.0)(7, 0.0)(8, 0.0)(9, 0.0)(10, 0.0)(11, 0.0)(12, 0.0)(13, 0.0)(14, 0.0)(15, 0.0)(16, 0.0)(17, 0.0)(18, 0.0)(19, 0.0)(20, 0.0)
    };

\addplot[
    color=blue,
    mark=square,mark options={scale=0.5}
    ]
    coordinates {
    (1, 0.0)(2, 0.0)(3, 0.019495853368062206)(4, 0.038970549768079475)(5, -0.5844254510063495)(6, -2.2988946562951034)(7, -2.493532239606728)(8, -2.4348321899539145)(9, -1.8116981843608506)(10, -1.148696004908814)(11, -0.797831857329584)(12, -0.057686271298629244)(13, 0.7409221220201434)(14, 1.480832723544359)(15, 2.045569650611989)(16, 2.2399348693191765)(17, 2.1037988113831396)(18, 2.2400442500077333)(19, 2.2986069884291913)(20, 2.162299416890545)
    };

    \end{groupplot}
    \node at ($(group c2r1) + (0cm,-3.8cm)$) {\ref{grouplegend}}; 
\end{tikzpicture}
    \caption{Relative success rate w.r.t different turns.}
    \label{fig:turn}
\end{figure}

%% file: fig/patience.tex
\begin{figure}[t]
    \centering
    \begin{tikzpicture}
    \pgfplotsset{footnotesize,samples=10}
    \begin{groupplot}[group style = {group size = 2 by 2, horizontal sep = 20pt , vertical sep = 25pt}, width = 6.4cm, height = 4.0cm]

\nextgroupplot[scaled ticks=false, tick label style={/pgf/number format/fixed}, title = {Computer Science},
legend style = { column sep = 10pt, legend columns = -1, legend to name = grouplegend2}, 
    ylabel={Success Rate \%}, 
            xlabel={Patience Threshold},
    xmin=0, xmax=6,
    ymin=0, ymax=50,
    xtick={1,2,3,4,5},
    ytick={0,10,20,30,40,50}, grid=both,
    grid style={dashed, gray!50},]

\addplot[
    color=brown, mark=diamond,
    ]
    coordinates {
    (1,0)(2,0.9)(3,7.2)(4,27.3)(5,44)
    };

\addplot[
    color=purple, mark=x,
    ]
    coordinates {
    (1,0)(2,1.8)(3,9.6)(4,27.3)(5,47)
    };

\addplot[
    color=red,
    mark=o,
    ]
    coordinates {
    (1,11.4)(2,16.3)(3,24.6)(4,31.4)(5,42.8)
    };

\addplot[
    color=blue,
    mark=square,
    ]
    coordinates {
    (1,11.4)(2,20.7)(3,31.3)(4,37.5)(5,43.1)
    };

    \addlegendentry{KNN}
    \addlegendentry{Greedy}
    \addlegendentry{PDDDQN}
    \addlegendentry{\texttt{PAI}}
    
\nextgroupplot[scaled ticks=false, tick label style={/pgf/number format/fixed}, title = {Math},
    xlabel={Patience Threshold},
    xmin=0, xmax=6,
    ymin=0, ymax=70,
    xtick={1,2,3,4,5},
    ytick={10,30,50,70}, grid=both,
    grid style={dashed, gray!50},]

\addplot[
    color=brown, mark=diamond,
    ]
    coordinates {
    (1,0)(2,6)(3,22.4)(4,46.9)(5,67.2)
    };

\addplot[
    color=purple, mark=x,
    ]
    coordinates {
    (1,0)(2,6.6)(3,25.6)(4,49.1)(5,69.8)
    };

\addplot[
    color=red,
    mark=o,
    ]
    coordinates {
    (1,21.1)(2,31.8)(3,40.1)(4,48.8)(5,53.8)
    };

\addplot[
    color=blue,
    mark=square,
    ]
    coordinates {
    (1,19)(2,37)(3,47.4)(4,53.4)(5,57.2)
    };
    
\nextgroupplot[scaled ticks=false, tick label style={/pgf/number format/fixed},ylabel={Impatience},
            xlabel={Patience Threshold},
    xmin=0, xmax=6,
    ymin=1, ymax=5,
    xtick={1,2,3,4,5},
    ytick={1,2,3,4,5}, grid=both,
    grid style={dashed, gray!50},]

\addplot[
    color=brown, mark=diamond,
    ]
    coordinates {
    (1,1.328)(2,2.390)(3,3.344)(4,4.118)(5,4.780)
    };

\addplot[
    color=purple, mark=x,
    ]
    coordinates {
    (1,1.339)(2,2.388)(3,3.325)(4,4.052)(5,4.681)
    };

\addplot[
    color=red,
    mark=o,
    ]
    coordinates {
    (1,0.876)(2,1.481)(3,1.991)(4,2.397)(5,2.745)
    };

\addplot[
    color=blue,
    mark=square,
    ]
    coordinates {
    (1,1.206)(2,1.852)(3,2.384)(4,2.851)(5,3.249)
    };

\nextgroupplot[scaled ticks=false, tick label style={/pgf/number format/fixed},
            xlabel={Patience Threshold},
    xmin=0, xmax=6,
    ymin=1, ymax=5,
    xtick={1,2,3,4,5},
    ytick={1,2,3,4,5}, grid=both,
    grid style={dashed, gray!50},]

\addplot[
    color=brown, mark=diamond,
    ]
    coordinates {
    (1,1.218)(2,2.228)(3,3.064)(4,3.710)(5,4.167)
    };

\addplot[
    color=purple, mark=x,
    ]
    coordinates {
    (1,1.293)(2,2.290)(3,3.081)(4,3.696)(5,4.111)
    };

\addplot[
    color=red,
    mark=o,
    ]
    coordinates {
    (1,1.002)(2,1.569)(3,2.010)(4,2.323)(5,2.607)
    };

\addplot[
    color=blue,
    mark=square,
    ]
    coordinates {
    (1,1.069)(2,1.586)(3,1.920)(4,2.131)(5,2.296)
    };

    \end{groupplot}
    \node at ($(group c2r2) + (-2.5cm,-2.8cm)$) {\ref{grouplegend2}}; 
\end{tikzpicture}
    \caption{Comparisons of success rates$\uparrow$ and impatience scores$\downarrow$ w.r.t students with different patience thresholds. 
%Full results are presented in Appendix \ref{app:detail_results}. 
}
    \label{fig:patience}
\end{figure}

%% file: fig/learn_rate.tex
\begin{figure}[t]
    \centering
    \begin{tikzpicture}
    \pgfplotsset{footnotesize,samples=10}
    \begin{groupplot}[group style = {group size = 2 by 2, horizontal sep = 20pt, vertical sep = 25pt}, width = 6.4cm, height = 4.0cm]

\nextgroupplot[scaled ticks=false, tick label style={/pgf/number format/fixed}, title = {Computer Science},
legend style = { column sep = 10pt, legend columns = -1, legend to name = grouplegend2}, 
    ylabel={Success Rate \%}, 
            xlabel={Learning Rate},
    xmin=0, xmax=0.06,
    ymin=0, ymax=80,
    xtick={0.01,0.02,0.03,0.04,0.05},
    ytick={0,20,40,60,80}, grid=both,
    grid style={dashed, gray!50},]

\addplot[
    color=brown, mark=diamond,
    ]
    coordinates {
    (0.01,2.1)(0.02,27.6)(0.03,52.2)(0.04,70.4)(0.05,76.5)
    };

\addplot[
    color=purple, mark=x,
    ]
    coordinates {
    (0.01,2.4)(0.02,27.3)(0.03,54.9)(0.04,80.0)(0.05,32.4)
    };

\addplot[
    color=red,
    mark=o,
    ]
    coordinates {
    (0.01,9.8)(0.02,31.4)(0.03,48.2)(0.04,53.9)(0.05,55.1)
    };

\addplot[
    color=blue,
    mark=square,
    ]
    coordinates {
    (0.01,9.1)(0.02,37.5)(0.03,49.6)(0.04,55.2)(0.05,57.5)
    };

    \addlegendentry{KNN}
    \addlegendentry{Greedy}
    \addlegendentry{PDDDQN}
    \addlegendentry{\texttt{PAI}}
    
\nextgroupplot[scaled ticks=false, tick label style={/pgf/number format/fixed}, title = {Math},
    xlabel={Learning Rate},
    xmin=0, xmax=0.06,
    ymin=0, ymax=90,
    xtick={0.01,0.02,0.03,0.04,0.05},
    ytick={0,20,40,60,80}, grid=both,
    grid style={dashed, gray!50},]

\addplot[
    color=brown, mark=diamond,
    ]
    coordinates {
    (0.01,7.3)(0.02,46.9)(0.03,72.5)(0.04,83.9)(0.05,87.4)
    };

\addplot[
    color=purple, mark=x,
    ]
    coordinates {
    (0.01,10.6)(0.02,49.1)(0.03,73.3)(0.04,89.9)(0.05,44.0)
    };

\addplot[
    color=red,
    mark=o,
    ]
    coordinates {
    (0.01,18.5)(0.02,48.8)(0.03,59.6)(0.04,63.6)(0.05,63.7)
    };

\addplot[
    color=blue,
    mark=square,
    ]
    coordinates {
    (0.01,29.3)(0.02,53.4)(0.03,60.5)(0.04,63.0)(0.05,64.1)
    };
    
\nextgroupplot[scaled ticks=false, tick label style={/pgf/number format/fixed},ylabel={Impatience},
            xlabel={Learning Rate},
    xmin=0, xmax=0.06,
    ymin=1, ymax=5,
    xtick={0.01,0.02,0.03,0.04,0.05},
    ytick={1,2,3,4,5}, grid=both,
    grid style={dashed, gray!50},]

\addplot[
    color=brown, mark=diamond,
    ]
    coordinates {
    (0.01,4.436)(0.02,4.118)(0.03,3.665)(0.04,3.276)(0.05,2.974)
    };

\addplot[
    color=purple, mark=x,
    ]
    coordinates {
    (0.01,4.427)(0.02,4.052)(0.03,3.628)(0.04,2.902)(0.05,2.006)
    };

\addplot[
    color=red,
    mark=o,
    ]
    coordinates {
    (0.01,3.256)(0.02,2.397)(0.03,1.851)(0.04,1.514)(0.05,1.300)
    };

\addplot[
    color=blue,
    mark=square,
    ]
    coordinates {
    (0.01,3.865)(0.02,2.851)(0.03,2.183)(0.04,1.702)(0.05,1.552)
    };

\nextgroupplot[scaled ticks=false, tick label style={/pgf/number format/fixed},
            xlabel={Learning Rate},
    xmin=0, xmax=0.06,
    ymin=1, ymax=5,
    xtick={0.01,0.02,0.03,0.04,0.05},
    ytick={1,2,3,4,5}, grid=both,
    grid style={dashed, gray!50},]

\addplot[
    color=brown, mark=diamond,
    ]
    coordinates {
    (0.01,4.302)(0.02,3.71)(0.03,3.170)(0.04,2.751)(0.05,2.457)
    };

\addplot[
    color=purple, mark=x,
    ]
    coordinates {
    (0.01,4.295)(0.02,3.696)(0.03,3.166)(0.04,2.551)(0.05,1.812)
    };

\addplot[
    color=red,
    mark=o,
    ]
    coordinates {
    (0.01,3.659)(0.02,2.323)(0.03,1.616)(0.04,1.242)(0.05,1.124)
    };

\addplot[
    color=blue,
    mark=square,
    ]
    coordinates {
    (0.01,3.277)(0.02,2.131)(0.03,1.670)(0.04,1.396)(0.05,1.275)
    };

    \end{groupplot}
    \node at ($(group c2r2) + (-2.5cm,-2.8cm)$) {\ref{grouplegend2}}; 
\end{tikzpicture}
    \caption{Comparisons of success rates$\uparrow$ and impatience scores$\downarrow$ w.r.t students with different learning rates. %Full results are presented in Appendix \ref{app:detail_results}. 
    }
    \label{fig:learn_rate}
\end{figure}